%% file: neurips.tex
\documentclass[preprint]{article}

\PassOptionsToPackage{sort, numbers, compress}{natbib}

\usepackage{neurips_2021}

\usepackage[utf8]{inputenc} 
\usepackage[T1]{fontenc}    
\usepackage{hyperref}       
\usepackage{url}            
\usepackage{booktabs}       
\usepackage{amsfonts}       
\usepackage{nicefrac}       
\usepackage{microtype}      
\usepackage{xcolor}         

\title{Asynchronous Distributed Optimization with Redundancy in Cost Functions}

\author{%
  Shuo Liu \\
  Georgetown University\\
  Washington, D.C., USA \\
  \texttt{sl1539@georgetown.edu} \\
   \And
   Nirupam Gupta \\
   Polytechnique Fédérale de Lausanne (EPFL). \\
   Lausanne, Switzerland \\
   \texttt{nirupam.gupta@epfl.ch} \\
   \AND
  Nitin H. Vaidya \\
  Georgetown University\\
  Washington, D.C., USA \\
  \texttt{nitin.vaidya@georgetown.edu} \\
}

\usepackage{amsmath,amsthm,amssymb}
\usepackage{tikz}
\usepackage{enumitem}
\usepackage[ruled,vlined]{algorithm2e}
\usepackage{multicol}

\tikzstyle{dot}=[circle,fill,inner sep=1.5pt]

\newtheorem{assumption}{Assumption}
\newtheorem{definition}{Definition}
\newtheorem{theorem}{Theorem}

\newtheorem{lemma}{Lemma}

\providecommand{\iprod}[2]{\ensuremath{\left\langle #1,\,#2  \right\rangle}}
\providecommand{\norm}[1]{\ensuremath{\left\lVert#1\right\rVert }}
\providecommand{\mnorm}[1]{\ensuremath{\left\lvert#1\right\rvert}}
\providecommand{\dist}[2]{\ensuremath{\mathrm{dist}\left( #1,\,#2 \right)}}

\def\R{\mathbb{R}}
\def\Z{\mathbb{Z}}
\def\H{\mathcal{H}}
\def\B{\mathcal{B}}

\def\D{\mathsf{D}}

\def\M{\mathsf{M}}
\def\W{\mathcal{W}}
\def\E{\mathbb{E}}
\def\g{\boldsymbol{g}}

\def\gf{\mathsf{GradFilter}}

\begin{document}

\maketitle

\begin{abstract}
    This paper considers the problem of asynchronous distributed multi-agent optimization on server-based system architecture. In this problem, each agent has a local cost, and the goal for the agents is to collectively find a minimum of their aggregate cost. A standard algorithm to solve this problem is the iterative distributed gradient-descent (DGD) method being implemented collaboratively by the server and the agents. In the synchronous setting, the algorithm proceeds from one iteration to the next only after all the agents complete their expected communication with the server. However, such synchrony can be expensive and even infeasible in real-world applications.
    We show that waiting for {\em all} the agents is unnecessary in many applications of distributed optimization, including distributed machine learning, due to {\em redundancy} in the cost functions (or {\em data}). Specifically, we consider a generic notion of redundancy named {\em $(r,\epsilon)$-redundancy} implying solvability of the original multi-agent optimization problem with $\epsilon$ accuracy, despite the removal of up to $r$ (out of total $n$) agents from the system. We present an asynchronous DGD algorithm where in each iteration the server only waits for (any) $n-r$ agents, instead of all the $n$ agents. Assuming {$(r,\epsilon)$-redundancy}, we show that our asynchronous algorithm converges to an approximate solution with error that is linear in $\epsilon$ and $r$. Moreover, we also present a generalization of our algorithm to tolerate some {\em Byzantine} faulty agents in the system. 
    Finally, we demonstrate the improved communication efficiency of our algorithm through experiments on MNIST and Fashion-MNIST using the benchmark neural network LeNet.
    
\end{abstract}

\input{intro}
\input{related}

\input{asynchrony_v2}
\input{extensions}

\input{experiment-learning}
\input{summary}

\section*{Broader Impact}



The main results of this paper are theoretical. Studied in this paper, the relationship between redundancy in cost functions and asynchronous distributed optimization solvability provides a framework that distributed optimization systems can generally adopt, such that robustness of such systems can be improved and theoretically backed. Since such a framework can be applied to various sorts of distributed optimization systems -- including distributed machine learning systems, the major ethical implications of our work root on the systems it would be applied to. Our work on asynchronous optimization can be used to boost robustness of some distributed systems with high failure rate, while the work on combination of asynchronous and fault-tolerant optimization can boost robustness of some distributed systems against failures and / or adversarial behaviors, which should be able to help improve the performance of distributed systems with untrustworthy participants.

\begin{ack}
Research reported in this paper was supported in part by the Army Research Laboratory under Cooperative Agreement W911NF- 17-2-0196, and by the National Science Foundation award 1842198. The views and conclusions contained in this document are those of the authors and should not be interpreted as representing the official policies, either expressed or implied, of the Army Research Laboratory, National Science Foundation or the U.S. Government. Research reported in this paper is also supported in part by a Fritz Fellowship from Georgetown University.
\end{ack}

\bibliographystyle{plain}
\begin{small}
    \bibliography{bib}
\end{small}

\appendix
\include{appendix-lemma}
\include{appendix-main}
\include{appendix-rate}
\include{appendix-gen}
\include{appendix-cge}

\end{document}

%% file: intro.tex
\section{Introduction}
\label{sec:intro}

With the rapid growth in both capability of modern computer systems and scale of optimization tasks, the problem of distributed optimization with multi-agent system has gained increasing attention in recent years. In the setting of multi-agent system, 
each agent has a local cost function. The goal is to design an algorithm that allows the agents to collectively minimize the aggregated cost functions of all agents. 
Formally, we suppose that there are $n$ agents in the system, and let $Q_i(x)$ denote the local cost function of agent $i$, where $x$ is a $d$-dimensional vector of real values, i.e., $x\in\R^d$. A distributed optimization algorithm outputs a global minimum $x^*$ such that 
\begin{equation}
    \textstyle x^*\in\arg\min_{x\in\R^d}\sum_{i=1}^nQ_i(x). \label{eqn:opt}
\end{equation}
For example, among a group of $n$ persons, the function $Q_i(x)$ can represent the cost of person $i$ to travel to some location $x$, and $x^*$ would be a location that minimizes the travel cost of all $n$ persons. The above multi-agent optimization problem finds many applications, including but not limited to distributed sensing \cite{rabbat2004distributed}, distributed learning \cite{boyd2011distributed}, swarm robotics \cite{raffard2004distributed}. 

Distributed optimization algorithms can be classified into \textit{synchronous} and \textit{asynchronous} algorithms. Synchronous algorithms require synchronous collective communication among the agents: communication in iterations, broadcasting and collecting vectors to and from agents. Such algorithms are widely adopted in many distributed computing packages \cite{dean2004mapreduce, zaharia2016apache} and distributed machine learning practice \cite{lin2014large}. However, synchronization of the agents is expensive in efficiency. For instance, in each iteration, agents with faster computation capability, better network connectivity, or less computing tasks would have to remain idle while waiting for slower agents (aka. \textit{stragglers}) to finish their task. 

In recent years, several approaches have been proposed to improve aforementioned efficiency cost. Generally speaking, the communication process is made asynchronous: instead of insisting on all agents complete their communication in the same iteration, computations are now allowed even with an incomplete iteration, vectors from different iterations (i.e., \textit{stale} vectors from stragglers), or both, with certain standard to maintain accuracy of the result or acceptable convergence rate \cite{iutzeler2013asynchronous, srivastava2011distributed, zhang2014asynchronous}.  

{\bf Redundancy in cost functions:} In many applications of distributed optimizations, including distributed sensing, distributed learning, and distributed linear regression, the agents' cost functions have some \textit{redundancy}.
Specifically, in a recent work, Liu et al.~defined a notion of $(2f,\epsilon)$-redundancy in distributed optimization and showed its necessity in tackling of Byzantine (adversarial) corruption of up to $f$ agents in the system~\cite{liu2021approximate}. Similar results exist in the context of resilient distributed learning~\cite{diakonikolas2018sever, diakonikolas2019robust}, and sensing~\cite{mao2019secure, mishra2016secure, shoukry2017secure}. We adopt such notions of redundancy for addressing the problem of asynchrony. Specifically, inspired from~\cite{gupta2020fault, liu2021approximate}, we consider the notion of $(r,\epsilon)$-redundancy.

Recall that the distance between a point $x$ and a set $Y$, denoted by $\dist{x}{Y}$, is defined to
    $\textstyle\dist{x}{Y}=\inf_{y\in Y}\dist{x}{y}=\inf_{y\in Y}\norm{x-y}$,
where $\norm{\cdot}$ represents Euclidean norm. The Hausdorff distance between two sets $X$ and $Y$ in $\mathbb{R}^d$, denoted by $\dist{X}{Y}$, is defined to be 
\begin{equation*}
    \textstyle\dist{X}{Y}\triangleq\max\left\{\sup_{x\in X}\dist{x}{Y}, \sup_{y\in Y}\dist{y}{X}\right\}.
\end{equation*}


\begin{definition}[$(r,\epsilon)$-redundancy]
\label{def:approx_redundancy}
For $r < n$ and $\epsilon \geq 0$, the agents' cost functions are said to satisfy $(r,\epsilon)$-redundancy if and only if a minimum of the aggregated of any $n-f$ (or more) costs is $\epsilon$-close to a minimum of the aggregate of all the $n$ costs. Specifically, for all $S \subseteq \{1, \ldots, \, n\}$ with $\mnorm{S}\geq n-r$, 
\begin{equation}
    \textstyle\dist{\arg\min_x\sum_{i\in S}Q_i(x)}{\arg\min_x\sum_{i\in[n]}Q_i(x)}\leq\epsilon.
\end{equation}
\end{definition}

In our context, we refer to $r$ as the {\em asynchrony parameter}, and $\epsilon$ as the {\em approximation parameter}. 

\begin{figure}[tb]
    \centering
    \begin{minipage}{.34\textwidth}
        \centering
        \resizebox{\textwidth}{!}{%
        \begin{tikzpicture}
            \draw (0,0)--(-3,-1.7);
            \draw (0,0)--(-1,-2);
            \draw (0,0)--(1,-2);
            \draw (0,0)--(3,-1.7);
            \node at (-3.5,-1) {\textit{agents}};
            \draw[fill=white] (0,0) ellipse (1 and .5) node[align=center]{\textit{server}};
            \draw[fill=white] (-3,-1.7) ellipse (.5 and .4) node[align=center]{1}; \node at (-2.3,-2.1) {$Q_1$};
            \draw[fill=white] (-1,-2) ellipse (.5 and .4) node[align=center]{2}; \node at (-.3,-2.4) {$Q_2$};
            \draw[fill=white] (1,-2) ellipse (.5 and .4) node[align=center]{3}; \node at (1.7,-2.4) {$Q_3$};
            \draw[fill=white] (3,-1.7) ellipse (.5 and .4) node[align=center]{4}; \node at (3.7,-2.1) {$Q_4$};
        \end{tikzpicture}
        }
        \caption{\footnotesize{System architecture.}}
        \label{fig:server-architecure}
    \end{minipage}\hfill
    \begin{minipage}{.65\textwidth}
        \centering
        \includegraphics[width=.63\textwidth]{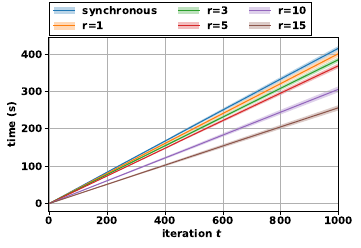}
        \caption{\footnotesize{Cumulative communication times for Algorithm~\ref{alg} with different $r$ values on Fashion-MNIST ($n=20$). See Section~\ref{sec:exp-learning} for details.}}
        \label{fig:mnist-time}
    \end{minipage}
\end{figure}

\paragraph{Our contributions:} We study the impact of the above redundancy property on the performance of asynchronous distributed optimization, i.e., its resilience to {\em stragglers}, in a server-based system architecture shown in Figure~\ref{fig:server-architecure}. Specifically, we consider an asynchronous distributed gradient-descent algorithm where in each iteration, unlike the synchronous setting, the server only waits for gradients from $n-r$ agents, instead of all the agents. For each agent $i$, $g^t_i$ denotes its gradient $\nabla Q_i(x^t)$ in iteration $t$. To ensure boundedness of estimates $x^t$, we constrain them to a convex compact set $\W$. 




\begin{algorithm}[H]
    \SetAlgoLined
    \caption{Asynchronous distributed gradient-descent with $(r,\epsilon)$-redundancy}
    \label{alg}
    The server chooses an arbitrary initial estimate $x^0$, updated in iteration  $t=0,1,2,\dots$ as follows:
    
    
\begin{description}[nosep]
    \item[Step 1:] The server requests each agent for the gradient of its local cost function at current estimate $x^t$. Each agent $j$ is expected to send to the server the gradient $\nabla Q_j(x^t)$ with timestamp $t$.  
    
    \item[Step 2:] The server waits until it receives $n-r$ gradients with the timestamp of $t$. Suppose $S^t\subsetneq\{1,...,n\}$ is the set of agents whose gradients are received by the server at step $t$ where $\mnorm{S^t}=n-r$. The server updates its estimate to
    \begin{equation}
        \textstyle x^{t+1}=\left[x^t-\eta_t \, \sum_{j\in S^t}\nabla Q_j(x^t) \right]_\W \label{eqn:update}
    \end{equation}
    where $\eta_t\geq0$ is the step-size for each iteration $t$.
\end{description}
\end{algorithm}


In Section~\ref{sec:redundancy}, we characterize the convergence of Algorithm~\ref{alg} under $(r,\epsilon)$-redundancy. 
Particularly, we show that our algorithm has {\em linearly} convergence rate when the aggregate cost is strongly convex, which is similar to the synchronous DGD; yet our algorithm has a lower net communication cost compared to the synchronous DGD, see Figure~\ref{fig:mnist-time}. We investigate the efficacy of our algorithm through experiments on benchmark machine learning problems of MNIST and Fashion-MNIST presented in Section~\ref{sec:exp-learning}. Empirical results show that we can obtain similar level of accuracy whilst saving on the communication cost by implementing Algorithm~\ref{alg}. We also present modifications to Algorithm~\ref{alg} to (i) accept stale gradients from previous iterations for further improved efficiency (see Section~\ref{sub:old-gradients}), and (ii) achieve Byzantine fault-tolerance in the asynchronous setting (see Section~\ref{sec:async-ft}).

\subsection{Applicability to distributed learning}
\label{sub:learning}

In the particular case of \textit{distributed learning}, each agent has some local \textit{data points} and the goal for the agent is to compute a parameter vector that best models the collective data points observed by all the agents \cite{bottou2018optimization}. Specifically, given a parameter vector $x$, for each data point $z$ we define a loss function $\ell(x;z)$. 
Suppose that the dataset possessed by agent $j$ is $\mathcal{D}_j$. The cost function of agent $i$ can be defined as 
\begin{equation}
     \textstyle Q_j(x)\triangleq\cfrac{1}{\mnorm{\mathcal{D}_j}}\sum_{z\in\mathcal{D}_j}\ell(x;z),
\end{equation}
where $\mnorm{\mathcal{D}_j}$ denotes the number of data points in dataset $\mathcal{D}_j$. 
The $(r,\epsilon)$-redundancy provides a trade-off between redundancy and the accuracy of the algorithm output. Our results on the benefits of redundancy in asynchronous optimization can also be extended for asynchronous distributed learning, where gradients of each agent $j$ can be computed by $\nabla Q_j(x)\triangleq\left(1/\mnorm{\mathcal{D}_j}\right)\sum_{z\in\mathcal{D}_j}\nabla\ell(x;z)$.

Furthermore, in practice, distributed machine learning algorithms often rely on stochastic gradients~\cite{bottou1998online,goyal2017accurate}, instead of full gradients. Specifically, in each iteration $t$, 
each agent $j$ randomly chooses $b$ data points $\boldsymbol{z}_j^t=\{z_{j_1}^t,...,z_{j_b}^t\}$ from dataset $\mathcal{D}_j$, where $b$ is often called the \textit{batch size}. The stochastic gradient is computed by
\begin{equation}
    \textstyle g_j^t=\cfrac{1}{b}\sum_{i=1}^b\nabla\ell\left(x^t;z_{j_i}^t\right).
\end{equation}
In expectation, $g_j^t$ equals the full gradient $Q_j(x)$, since $\E_{\boldsymbol{z}_j^t}({1}/{b})\sum_{i=1}^b\ell(x;z_{j_i}^t)=Q_j(x)$. 

Based on the above analysis, although our theoretical results in this paper are for full gradients, the empirical studies in Section~\ref{sec:exp-learning} use distributed stochastic gradient descent (D-SGD).

%% file: related.tex
\section{Related work}
\label{sec:related}


\textit{Stragglers}, or the agents that operate slowly, often dominates the running time of a distributed algorithm. Asynchronous algorithms that mitigate the effect of stragglers have been explored in both server-based and peer-to-peer system architectures. 

In a shared-memory system, where the clients and server communicate via shared memory, prior work has investigated asynchronous distributed optimization algorithms.
In particular, prior work shows that the distributed optimization problem can be solved using \textit{stale} gradients with a constant delay  \cite{langford2009slow} or with bounded delay \cite{agarwal2012distributed, feyzmahdavian2016asynchronous}.
Furthermore, methods such as \textsc{Hogwild!} allow lock-free update in shared memory \cite{niu2011hogwild}. There is also work that uses variance reduction and incremental aggregation methods
to improve the convergence rate \cite{roux2012stochastic, johnson2013accelerating, defazio2014saga, shalev2013accelerated, blatt2007convergent}. Readers can further refer to \cite{assran2020advances} for a detailed survey. 
Although in a different system architecture, these research indicate it is possible to solve asynchronous optimization problems using stale gradients. Still, a bound on allowable delay is needed, and convergence rate is related to that bound.

\textit{Coding} has been used to mitigate the effect of stragglers or failures \cite{tandon2017gradient, lee2017speeding, karakus2017encoded, karakus2017straggler, halbawi2018improving, yang2017coded, karakus2019redundancy}.  Tandon et al. \cite{tandon2017gradient} proposed a framework using maximum-distance separable coding across gradients to tolerate failures and stragglers. Similarly, Halbawi et al. \cite{halbawi2018improving} adopted coding to construct a coding scheme with a time-efficient online decoder.
Karakus et al. \cite{karakus2019redundancy} proposed an encoding distributed optimization framework with deterministic convergence guarantee.
Kadhe et al. \cite{kadhe2019gradient} considered stragglers in an adversarial setting, providing a way to construct approximate gradient codes lower-bounding the number of adversarial stragglers needed to inflict certain approximation error. Baharav et al. \cite{baharav2018straggler} presented a product-code-based method for distributed matrix multiplication with stragglers.
Other replication- or repetition-based techniques involve either task-rescheduling or assigning the same tasks to multiple nodes \cite{ananthanarayanan2013effective, gardner2015reducing, shah2015redundant, wang2015using, yadwadkar2016multi}.
These previous methods rely on algorithm-created redundancy of data or gradients to achieve robustness. The $(r,\epsilon)$-redundancy, however, is a property of the cost functions themselves. Our Algorithm~\ref{alg} exploits such redundancy without extra effort. 


Our work is mainly inspired by past work in Byzantine fault-tolerant distributed optimization showing necessity of {\it redundancy} in the cost functions to tolerate Byzantine failures~\cite{gupta2020fault, liu2021approximate}. 
Byzantine fault-tolerant algorithms in distributed optimization aim to output a minimum point of the aggregated cost functions of only non-faulty agents in the system in presence of up to $f$ faulty agents \cite{su2016fault}, while making no assumptions on the information sent by \textit{Byzantine} faulty agents \cite{lamport2019byzantine}. Formally, suppose among $n$ agents, a subset $\B$ of agents are Byzantine faulty ($\mnorm{\B}\leq f$) while the reminder $\H$ are non-faulty ($\{1,\dots,n\}\backslash\B=\H$). The goal is to find a point $\widehat{x}$ such that
\begin{equation}
    \label{eqn:ft-goal}
    \textstyle\widehat{x}\in\arg\min_{x}\sum_{j\in\H}Q_j(x).
\end{equation}
A variety of approaches have been proposed to address this problem in different ways. 

In particular, Liu et al. \cite{liu2021approximate} has shown that the agents' cost functions must satisfy the property of \textit{$(2f,\epsilon)$-redundancy}, 
to solve the distributed optimization problem 
with $\epsilon$ accuracy with up to $f$ Byzantine faulty agents.
In the special case when $\epsilon=0$, $(2f,0)$-redundancy (or simply \textit{$2f$-redundancy}) is both necessary and sufficient to solve the Byzantine fault-tolerant distributed optimization problem \textit{exactly}~\cite{gupta2020fault}. These work indicates that the redundancy, as a property of the cost functions, can be exploited to obtain robustness for a distributed optimization algorithm. 


The goal of asynchronous optimization studied in this paper is to mitigate the impact of stragglers. We exploit the redundancy of cost functions, similar to that used in the works for the purpose of fault-tolerance. As described formally later in Section~\ref{sec:redundancy}, Algorithm~\ref{alg} utilizes redundancy of cost functions to provide provable convergence properties. We further show that stale gradients can also be accepted, and the error of output does not depend on the bound on staleness. Also,
such redundancy can be used to achieve robustness against stragglers and Byzantine faulty agents at the same time. Our results are applicable to many applications of distributed optimizations, including distributed sensing~\cite{alouani2005theory, ji2012distributed}, distributed learning~\cite{grishchenko2018asynchronous, farina2019asynchronous, mcmahan2014delay}, and distributed linear regression~\cite{liu2021approximate}.

%% file: asynchrony_v2.tex
\section{Exploiting Redundancy for Asynchronous Distributed Optimization}
\label{sec:redundancy}



In this section, we consider the general \textit{$(r,\epsilon)$-redundancy}, introduced in Definition~\ref{def:approx_redundancy}, and provide convergence analysis of Algorithm~\ref{alg} given the cost functions of agents satisfies $(r,\epsilon)$-redundancy. 



From now on, we use $[n]$ as a shorthand for the set $\{1,...,n\}$. Suppose the cost function for agent $i$ is $Q_i(x)$. We first introduce some assumptions on the cost functions that are necessary for our analysis.

\begin{assumption}
    \label{assum:lipschitz}
    For each $i$, the function $Q_i(x)$ is $\mu$-Lipschitz smooth, i.e., $\forall x, x'\in\mathbb{R}^d$, 
    \begin{equation}
        \norm{\nabla Q_i(x)-\nabla Q_i(x')}\leq\mu\norm{x-x'}.
    \end{equation}
\end{assumption}

\begin{assumption}
    \label{assum:strongly-convex}
    For any set $S\subset[n]$, we define the average cost function to be $Q_S(x)=({1}/{\mnorm{S}})\sum_{j\in S}Q_j(x)$. We assume that $Q_S(x)$ is $\gamma$-strongly convex for any $S$ subject to $\mnorm{S}\geq n-r$, i.e., $\forall x, x'\in\mathbb{R}^d$, 
    \begin{equation}
        \iprod{\nabla Q_S(x)-\nabla Q_S(x')}{x-x'}\geq\gamma\norm{x-x'}^2.
    \end{equation}
\end{assumption}

Given Assumption~\ref{assum:strongly-convex}, there is only one minimum point for any $S$, $\mnorm{S}\geq n-r$. Let us define $ x^*=\arg\min_x\sum_{j\in[n]}Q_j(x)$.
Thus, $x^*$ is the unique minimum point for the aggregated cost functions of all the agents. We assume that (recall that $\W$ is used in (\ref{eqn:update}))
\begin{equation}
    \label{eqn:existence-exact}
    x^*\in \W.
\end{equation}

For the convenience of analysis, let us first define a generalized notation of gradient aggregation rule $\mathsf{GradAgg}:\R^{d\times n}\rightarrow\R^d$, a function mapping $n$ gradients to a $d$-dimensional vector. Under this notation, the iterative update of estimates in any aggregation-based algorithm can be written as
\begin{equation}
    x^{t+1}=\left[x^t-\eta_t\mathsf{GradAgg}\left(\nabla Q_1(x^t), ..., \nabla Q_n(x^t)\right)\right]_\W.
    \label{eqn:update-general}
\end{equation}
In our Algorithm~\ref{alg}, recall \eqref{eqn:update} in \textbf{Step 2}, the gradient aggregation rule is
\begin{equation}
    \textstyle\mathsf{GradAgg}\left(\nabla Q_1(x^t), ..., \nabla Q_n(x^t)\right)=\sum_{j\in{S^t}}\nabla Q_j(x^t)
    \label{eqn:aggregation-rule}
\end{equation}
for every iteration $t$. Note that although all $n$ gradients are formally mentioned in the aggregation rule notation, only those in $S^t$ are actually used, and in practice the algorithm only need to wait for $n-r$ gradients. With this notation, we state the lemma:

\begin{lemma}
    \label{lemma:bound}
    Consider the general iterative update rule (\ref{eqn:update-general}). Let $\eta_t$ satisfy $\sum_{t=0}^\infty\eta_t=\infty~\textrm{ and }~\sum_{t=0}^\infty\eta_t^2<\infty$.
    For any given gradient aggregation rule $\mathsf{GradAgg}\left(\nabla Q_1(x^t), ..., \nabla Q_n(x^t)\right)$, if there exists $\M<\infty$ such that
    \begin{equation}
        \norm{\mathsf{GradAgg}\left(\nabla Q_1(x^t), ..., \nabla Q_n(x^t)\right)}\leq\M
        \label{eqn:bounded-aggregator-norm}
    \end{equation} for all $t$, and there exists $\D^*\in\left[0,\max_{x\in\W}\norm{x-x^*}\right)$ and $\xi>0$ such that when $\norm{x^t-x^*}\geq\D^*$,
    \begin{equation}
        \phi_t\triangleq\iprod{x^t-x^*}{\mathsf{GradAgg}\left(\nabla Q_1(x^t), ..., \nabla Q_n(x^t)\right)}\geq\xi,~
        \label{eqn:phi-def}
    \end{equation}
    we have $\lim_{t\rightarrow\infty}\norm{x^t-x^*}\leq\D^*$.
\end{lemma}
Note that $\D^*$ and $\xi$ 
need not be independent. 
With Lemma~\ref{lemma:bound}, we present below an asymptotic convergence guarantee of our proposed algorithm described in Section~\ref{sec:intro}. 
\begin{theorem}
     Suppose that Assumptions~\ref{assum:lipschitz} and \ref{assum:strongly-convex} hold true, and the agents' cost functions satisfy $(r,\epsilon)$-redundancy. 
     Assume that $\eta_t$ in (\ref{eqn:update}) satisfies $\sum_{t=0}^\infty\eta_t=\infty$ and $\sum_{t=0}^\infty\eta_t^2<\infty$.
     Let $\alpha$ and $D$ be defined as follows: $\alpha\triangleq 1-\dfrac{r}{n}\cdot\dfrac{\mu}{\gamma}>0$ and $\D\triangleq\dfrac{2r\mu}{\alpha\gamma}\epsilon$. Then, for the proposed Algorithm~\ref{alg}, 
        $\lim_{t\rightarrow\infty}\norm{x^t-x^*}\leq\D$, 
    where $x^*=\arg\min_x\sum_{i\in[n]}Q_i(x)$.
    \label{thm:approx}
\end{theorem}
Intuitively, when $(r,\epsilon)$-redundancy is satisfied, our algorithm is guaranteed to output an approximation of the true minimum point of the aggregated cost functions of all agents, and the distance between algorithm's output and the true minimum $x^*=\arg\min_x\sum_{i\in[n]}Q_i(x)$ is bounded. The error bound $\D$ is linear to $r$ and $\epsilon$.

We obtain below the convergence rate of Algorithm~\ref{alg} for two different types of step-sizes. 
\begin{theorem}
    \label{thm:conv-rate}
    Suppose Assumptions~\ref{assum:lipschitz} and \ref{assum:strongly-convex} hold true, and the agents' cost functions satisfy $(r,\epsilon)$-redundancy. Define $\alpha=1-\cfrac{r}{n}\cdot\cfrac{\mu}{\gamma}$ and $\overline{\eta}=\cfrac{2\gamma\alpha}{\mu^2 n}$. 
    There exists a positive real-value $R \in \Theta(\epsilon)$, such that Algorithm~\ref{alg} satisfy (i) $A\in[0,1)$, and (ii) $\norm{x^t-x^*}^2\leq A^t\norm{x^0-x^*}+R$, if
    \begin{enumerate}[nosep,label=\alph*),leftmargin=*]
        \item let $A=1-(\mu n)^2\eta(\overline{\eta}-\eta)$ for some $\eta$, and $\eta_t=\eta$ for every iteration $t$ with $\eta\in(0,\overline{\eta})$.  \label{thm:conv-rate-a}
        \item let $A=1-(\mu n)^2c(\overline{\eta}-c)$ for some $c$, and $\eta_t=c/(t+1)$ for every iteration $t$ with $c\in(0,\overline{\eta})$. \label{thm:conv-rate-b}
    \end{enumerate}
    
\end{theorem}
According to Theorem~\ref{thm:conv-rate}, the sequence of estimates $\left\{x^t\right\}$ generated by our algorithm converges \textit{linearly} to a ball centered by the true minimum $x^*$ of radius $R$, if the step-size $\eta_t$ is a) a fixed value $\eta$, or b) a diminishing sequence $c/(t+1)$. 
Also, with $R=\Theta(\epsilon)$ indicating that $R=0$ when $\epsilon=0$, our result in  Theorem~\ref{thm:conv-rate}\ref{thm:conv-rate-a} generalizes the linear convergence rate result of gradient descent with constant step-size \cite{nesterov2003introductory}. 

\subsection{The Special case of exact redundancy}
\label{sec:exact-redundancy}

Now we consider a special case when the cost functions of agents satisfies $(r,0)$-redundancy. Note that $(r,0)$-redundancy 
is equivalent to the $r$-redundancy property defined as follows.

\begin{definition}[$r$-redundancy]
    \label{def:redundancy}
    For any two sets $S$ and $T$, each containing at least $n-r$ agents, 
    \begin{equation}
        \textstyle\arg\min_x\sum_{j\in S}Q_j(x)=\arg\min_x\sum_{j\in T}Q_j(x).
        \label{eqn:redundancy}
    \end{equation}
\end{definition}
\noindent Intuitively, $r$-redundancy indicates that all sets of more than $n-r$ agents have their aggregated cost functions minimize at the same set of points. The following is a corollary from $r$-redundancy.

\begin{lemma}
\label{lemma:minimum}
Under Assumptions~\ref{assum:lipschitz} and \ref{assum:strongly-convex}, $\nabla Q_j(x^*)=0$ for all $j\in[n]$, if the cost functions of all agents satisfies $r$-redundancy property.
\end{lemma}
With $r$-redundancy, our algorithm can output $x^*$, the minimum point of the aggregation of cost functions of all the agents. 

\begin{theorem}
    Suppose Assumptions~\ref{assum:lipschitz} and \ref{assum:strongly-convex} hold true, and the cost functions of all agents satisfies $r$-redundancy. Assume that $\eta_t$ in (\ref{eqn:update}) satisfies $\sum_{t=0}^\infty\eta_t=\infty$ and $\sum_{t=0}^\infty\eta_t^2<\infty$, Algorithm~\ref{alg} has 
        $\lim_{t\rightarrow\infty}x^t=x^*$, 
    where $x^*=\arg\min_x\sum_{i\in[n]}Q_i(x)$.
    \label{thm:exact}
\end{theorem}

\subsection{Improving efficiency using gradients from past iterations}
\label{sub:old-gradients}


Let us denote $T^{t;k}$ the set of agents whose latest gradients received by the server at iteration $t$ is computed using the estimate $x^{k}$. Suppose $S^{t;k}$ is the set of agents whose gradients computed using the estimate $x^t$ is received by iteration $t$. $T^{t;k}$ can be defined in an inductive way: (i) $T^{t;t}=S^{t;t}$, and (2) $T^{t;t-i}=S^{t;t-i}\backslash\bigcup_{k=0}^{i-1}T^{t;t-k}$, $\forall i\geq1$.
Note that the definition implies $T^{t;t-i}\cap T^{t;t-j}$ for any $i\neq j$. Let us further define $T^t=\bigcup_{i=0}^\tau T^{t;t-i}$, where $\tau\geq0$ is a predefined \textit{straggler parameter}.

To extend Algorithm~\ref{alg} to use gradients of previous iterations, in \textbf{Step 2} we replace update rule \eqref{eqn:update} to
\begin{equation}
    \label{eqn:update-straggler}
    \textstyle x^{t+1}=\left[x^t-\eta_t\sum_{i=0}^\tau\sum_{j\in T^{t;t-i}}\nabla Q_j(x^{t-i})\right]_\W,
\end{equation}
and now we only require the server wait till $T^t$ is at least $n-r$.
Intuitively, the new algorithm updates its iterative estimate $x^t$ using the latest gradients from no less than $n-r$ agents at each iteration $t$. 

\begin{theorem}
    Suppose Assumption~\ref{assum:lipschitz} and \ref{assum:strongly-convex} hold true, and the cost functions of all agents satisfies $(r,\epsilon)$-redundancy. 
     Assume that $\eta_t$ in (\ref{eqn:update-straggler}) satisfies $\sum_{t=0}^\infty\eta_t=\infty$, $\sum_{t=0}^\infty\eta_t^2<\infty$, and $\eta_t\geq\eta_{t+1}$ for all $t$.
     Let $\alpha$ and $D$ be defined as follows: 
     $\alpha\triangleq1-\dfrac{r}{n}\cdot\dfrac{\mu}{\gamma}>0 \textrm{ and } D\triangleq\dfrac{2r\mu}{\alpha\gamma}\epsilon$. 
     Then, suppose there exists a $\tau\geq0$ such that $\mnorm{T^t}\geq n-r$ for all $t$, for the proposed algorithm with update rule \eqref{eqn:update-straggler}, 
        $\lim_{t\rightarrow\infty}\norm{x^t-x^*}\leq\D$, 
    where $x^*=\arg\min_x\sum_{j\in[n]}Q_j(x)$.
    \label{thm:approx-generalized}
\end{theorem}
The new algorithm accepts gradients at most $\tau$-iteration stale. 
Theorem~\ref{thm:approx-generalized} shows exactly the same bound as is in Theorem~\ref{thm:approx}, independent from $\tau$, indicating that by using gradients from previous iterations, the accuracy of the output would not be effected, so long as the number of gradients used in each iteration is guaranteed by properly choosing $\tau$. Still, the convergence rate (number of iterations needed to converge) will be effected by $\tau$.

%% file: extensions.tex


\section{Asynchronous Byzantine fault-tolerant optimization}
\label{sec:async-ft}

The robustness provided by redundancy in cost functions against faulty agents and stragglers can be combined. 
Consider a Byzantine distributed optimization problem, where in an $n$-agent system, up to $f$ can be Byzantine faulty. As mentioned in Section~\ref{sec:related}, the goal \eqref{eqn:ft-goal} is to approximate a point $\widehat{x}$ that
\begin{equation}
    \textstyle\widehat{x}\in\arg\min_{x}\sum_{j\in\H}Q_j(x). \tag{\ref{eqn:ft-goal}}
\end{equation}
Also, we want our algorithm to be robust against up to $r$ stragglers, as we did above in Section~\ref{sec:redundancy}. Note that the sets for Byzantine agents and stragglers may or may not intersect. 

Correspondingly, we need to modify our redundancy property to include both Byzantine agents and stragglers. Specifically,  $(f,r;\epsilon)$-redundancy is defined as follows.
\begin{definition}[$(f,r;\epsilon)$-redundancy]
    \label{def:approx_redundancy-ft}
    The agents cost functions are said to satisfy $(f,r;\epsilon)$-redundancy, if and only if the distance between the minimum point sets of the aggregated cost functions of any pair of subsets of non-faulty agents $S,\widehat{S}\subsetneq\{1,...,n\}$, where $\mnorm{S}=n-f$, $\mnorm{\widehat{S}}\geq n-r-2f$ and $\widehat{S}\subsetneq S$, is bounded by $\epsilon$, i.e.,
    \begin{equation}
        \textstyle\dist{\arg\min_x\sum_{i\in S}Q_i(x)}{\arg\min_x\sum_{i\in\widehat{S}}Q_i(x)}\leq\epsilon,
    \end{equation}
\end{definition}
Following analysis in \cite{liu2021approximate}, we know that such redundancy is \textit{necessary} for solving a Byzantine fault-tolerant distributed optimization problem. 


Robust aggregation rules, or gradient filters, is used to achieve fault-tolerance in distributed optimization \cite{liu2021approximate}. A gradient filter $\mathsf{GradFilter}(m,f;\cdot):\R^{d\times m}\rightarrow\R^d$ is a function that takes $m$ vectors of $d$-dimension, and output a $d$-dimension vector given that there are up to $f$ Byzantine agents. Generally, $m>f\geq0$. Suppose each agent $i$ sends a vector
\begin{equation}
    g_i^t=\left\{\begin{array}{cl}
        \nabla Q_i(x^t), & \textrm{ if the agent is non-faulty,} \\
        \textrm{arbitrary vector}, & \textrm{ if the agent is faulty} 
    \end{array}\right.
\end{equation}
to the server at iteration $t$. Our Algorithm~\ref{alg} with the general iterative update \eqref{eqn:update-general} can be used to achieve fault-tolerance with asynchronous optimization, when 
\begin{equation}
    \mathsf{GradAgg}\left(\nabla Q_1(x^t), ..., \nabla Q_n(x^t)\right)=\mathsf{GradFilter}\left(n-r,f;\,\left\{g_i^t\right\}_{i\in S^t}\right).
    \label{eqn:aggregation-rule-ft}
\end{equation}
Following the above aggregation rule, the server receives the first $n-r$ vectors from the agents in the set $S^t$, and send the vectors through a gradient filter. 

\begin{theorem}
    \label{thm:async-fault-toler}
    Suppose that Assumptions~\ref{assum:lipschitz} and \ref{assum:strongly-convex} hold true, and the cost functions of all agents satisfies $(f,r;\epsilon)$-redundancy. Assume that $\eta_t$ 
    satisfies $\sum_{t=0}^\infty=\infty$ and $\sum_{t=0}^\infty\eta_t^2<\infty$. Suppose that $\norm{\mathsf{GradFilter}\left(n-r,f;\,\left\{g_i^t\right\}_{i\in S^t}\right)}<\infty$ for all $t$. The proposed algorithm with aggregation rule \eqref{eqn:aggregation-rule-ft} satisfies the following: For some point $x^*\in\W$, if there exists a real-valued constant $\D^*\in\left[0,\max_{x\in\W}\norm{x-x^*}\right)$ and $\xi>0$ such that for each iteration $t$,
        $\phi_t\triangleq\iprod{x^t-x^*}{\mathsf{GradFilter}\left(n-r,f;\,\left\{g_i^t\right\}_{i\in S^t}\right)}\geq\xi$ when $\norm{x^t-x^*}\geq\D^*$,
    we have $\lim_{t\rightarrow\infty}\norm{x^t-x^*}\leq\D^*$.
\end{theorem}
Theorem~\ref{thm:async-fault-toler} already stands combining \eqref{eqn:aggregation-rule-ft} and Lemma~\ref{lemma:bound}. Intuitively, so long as the gradient filter in use satisfies the desired properties in Theorem~\ref{thm:async-fault-toler}, our adapted algorithm can tolerate up to $f$ Byzantine faulty agents and up to $r$ stragglers in distributed optimization. Such kind of gradient filters includes CGE \cite{liu2021approximate} and coordinate-wise trimmed mean \cite{yin2018byzantine}. 
Here we present the detailed result for CGE.

To introduce the convergence property with a specific gradient filter, similarly, 
we also need some assumptions on the cost functions. Note that the following assumptions only apply to non-faulty agents. Suppose $\H\subset[n]$ is a subset of non-faulty agents with $\mnorm{\H}=n-f$. 
\begin{assumption}
    \label{assum:lipschitz-ft}
    For each non-faulty agent $i$, the function $Q_i(x)$ is $\mu$-Lipschitz smooth, i.e., $\forall x, x'\in\mathbb{R}^d$, 
    \begin{equation}
        \norm{\nabla Q_i(x)-\nabla Q_i(x')}\leq\mu\norm{x-x'}.
    \end{equation}
\end{assumption}

\begin{assumption}
    \label{assum:strongly-convex-ft}
    For any set $S\subset\H$, we define the average cost function to be $Q_S(x)=({1}/{\mnorm{S}})\sum_{j\in S}Q_j(x)$. We assume that $Q_S(x)$ is $\gamma$-strongly convex for any $S$ subject to $\mnorm{S}\geq n-f$, i.e., $\forall x, x'\in\mathbb{R}^d$, 
    \begin{equation}
        \iprod{\nabla Q_S(x)-\nabla Q_S(x')}{x-x'}\geq\gamma\norm{x-x'}^2.
    \end{equation}
\end{assumption}
Similar to the conclusion in Section~\ref{sec:redundancy} regarding the existence of a minimum point \eqref{eqn:existence-exact}, we also require the existence of a solution to the fault-tolerant optimization problem: 
    For each subset of non-faulty agents $S$ with $\mnorm{S}=n-f$, we assume that there exists a point $x_S\in\arg\min_{x\in\R^d}\sum_{j\in S}Q_j(x)$ such that $x_S\in\W$.
Suppose $\H$ is an arbitrary set of non-faulty agents with $\mnorm{\H}=n-r$. By Assumptions~\ref{assum:strongly-convex-ft}, 
there exists a unique minimum point $x_\H$ in $\W$ that minimize the aggregated cost functions of agents in $\H$. Specifically,
    $\{x_\H\}=\W\cap\arg\min_{x\in\R^d}\sum_{j\in\H}Q_j(x)$.

We state the convergence property when applying CGE as follows. 
\begin{theorem}
    \label{thm:async-fault-toler-cge}
    Suppose that Assumptions~\ref{assum:lipschitz} and \ref{assum:strongly-convex} 
    hold true, and the cost functions of all agents satisfies $(f,r;\epsilon)$-redundancy. Assume that $\eta_t$ 
    satisfies $\sum_{t=0}^\infty=\infty$ and $\sum_{t=0}^\infty\eta_t^2<\infty$. If the following conditions holds true: 
    (i) $\norm{\mathsf{GradFilter}\left(n-r,f;\,\left\{g_i^t\right\}_{i\in S^t}\right)}<\infty$ for all $t$, and 
    (ii) if 
            $\displaystyle\alpha = 1 - \frac{f-r}{n-r} + \frac{2\mu}{\gamma}\cdot\frac{f+r}{n-r}>0$,
        then for each set of $n-f$ non-faulty agents $\H$, for an arbitrary $\delta > 0$,
            $\displaystyle\phi_t \geq \alpha m \gamma \delta \left( \frac{4\mu (f+r)\epsilon}{\alpha\gamma} + \delta \right) > 0$ when $\displaystyle\norm{x^t-x_{\H}}\geq\frac{4\mu (f+r)\epsilon}{\alpha\gamma}+\delta$.
    Then for the proposed algorithm with aggregation rule \eqref{eqn:aggregation-rule-ft}, we have $\lim_{t\rightarrow\infty}\norm{x^t-x_\H}\leq\cfrac{4\mu (f+r)\epsilon}{\alpha\gamma}$.
\end{theorem}
Intuitively, by applying CGE gradient filter, the output of our new algorithm can converge to a $\epsilon$-related area centered by the true minimum point $x_\H$ of aggregated cost functions of non-faulty agents, with up to $f$ Byzantine agents and up to $r$ stragglers. 

Note that when $f=0$, $(f,r;\epsilon)$-redundancy becomes $(r,\epsilon)$-redundancy for asynchronous distributed optimization, while when $r=0$, the redundancy property becomes $(2f,\epsilon)$-redundancy \cite{liu2021approximate} for Byzantine fault-tolerant distributed optimization. In other words, $(f,r;\epsilon)$-redundancy is a generalized property for both asynchronous and fault-tolerant optimization\footnote{Theorem~\ref{thm:async-fault-toler-cge} does not match the special case of Theorem~\ref{thm:approx} when $f=0$, for some inequalities in the proofs are not strict bounds.}.

%% file: experiment-learning.tex
\section{Experiments}
\label{sec:exp-learning}

In this section, we present our empirical results for Algorithm~\ref{alg} on some benchmark distributed machine learning tasks, of which the applicability is discussed in Section~\ref{sub:learning}. It is worth noting that 
it is difficult to compute redundancy parameters of the cost functions in distributed learning tasks. Instead, we would like to empirically show the broader applicability of Algorithm~\ref{alg} and the existence of underlying redundancy which we can utilize in real-world tasks.

We simulate a server-based distributed learning system (ref. Figure~\ref{fig:server-architecure}) using multiple threads, one for the server and the rest for agents. The inter-thread communication is handled via message passing interface. The simulator is built in Python using PyTorch \cite{paszke2019pytorch} and MPI4py \cite{dalcin2011parallel}, and deployed on a Google Cloud Platform cluster with 14 vCPUs and 100 GB memory. 

The experiments are conducted on two benchmark datasets: MNIST and Fashion-MNIST. MNIST \cite{bottou1998online} is an image-classification dataset of monochrome handwritten digits. Fashion-MNIST \cite{xiao2017fashion} is an image-classification dataset of grayscale images of clothes. Both datasets comprise of 60,000 training and 10,000 testing data points, with each image of size $28\times28$. For each dataset, we train a benchmark neural network LeNet \cite{lecun1998gradient} which has 431,080 learnable parameters. 

\begin{figure}[t]
    \centering
    \includegraphics[width=.9\textwidth]{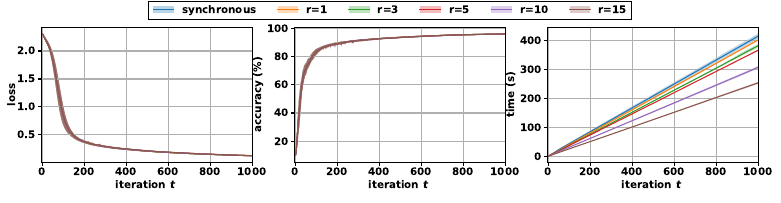}
    \caption{\textit{\small{Performance of Algorithm~\ref{alg} for MNIST using D-SGD. \emph{Synchronous} (in \emph{blue}) represents the typical synchronous distributed optimization algorithm (r=0) for comparison. The performance is presented by the \emph{losses} and \emph{accuracies}. Cumulative communication times over iterations are also presented. Experiment of each parameter setting is run 5 times and performance is averaged. The lines show the averaged performance, with shadows indicate standard deviations.}}}
    \label{fig:mnist}
\end{figure}

\begin{figure}[t]
    \centering
    \includegraphics[width=.9\textwidth]{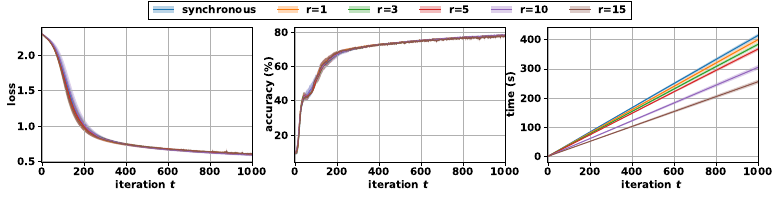}
    \caption{\textit{\small{Performance of Algorithm~\ref{alg} for Fashion-MNIST using D-SGD. \emph{Synchronous} (in \emph{blue}) represents the typical synchronous distributed optimization algorithm (r=0) for comparison. The performance is presented by the \emph{losses} and \emph{accuracies}. Cumulative communication times over iterations are also presented. Experiment of each parameter setting is run 5 times and performance is averaged. The lines show the averaged performance, with shadows indicate standard deviations.}}}
    \label{fig:fmnist}
\end{figure}

In each of our experiments, we simulate a distributed system of $n=20$ agents with different values of $r=0,1,3,5,10,15$. Note that when $r=0$, our algorithm becomes synchronous. We choose batch size $b=128$ for D-SGD, and fixed step-size $\eta=0.01$. The performance of our algorithm is measured by cross-entropy loss and model accuracy at each step. We also document the cumulative communication time of each setting. Experiments of each setting are run 5 times with different random seed\footnote{Randomnesses exist in drawing of data points and stragglers in each iteration of each execution.}, and the averaged performance is reported. The results are shown in Figure~\ref{fig:mnist} for MNIST and Figure~\ref{fig:fmnist} for Fashion-MNIST. We show the first 1,000 iterations as there is a clear trend of converging by the end of 1,000 iterations for both tasks in all four settings.

As is shown in the first two figures in Figures~\ref{fig:mnist} and \ref{fig:fmnist}, regardless of the asynchronous parameter $r$, Algorithm~\ref{alg} converges in a similar rate, and the learned model reaches comparable accuracy to the one learned by synchronous algorithm at the same iteration. However, by dropping out $r$ stragglers, the communication overhead is gradually reduced with increasing value of $r$. It is worth noting that 
reduction in communication time over $r$ when $r$ is small is more significant than that when $r$ is large, indicating a small number of stragglers work very slow, from which our algorithm gains more improvement in communication overhead.

%% file: summary.tex
\section{Summary}
\label{sec:summary}

We have studied the influence of redundancy in agents' cost functions on asynchronous distributed multi-agent optimization. Specifically, we have presented an asynchronous distributed optimization algorithm, and analyzed its convergence when agents' costs have $(r,\epsilon)$-redundancy - a generic characterization of redundancy in the costs. We have shown that, under $(r,\epsilon)$-redundancy, our Algorithm~\ref{alg} outputs an approximate solution of the distributed optimization problem with $\epsilon$ accuracy. We have also shown that redundancy in costs, when exploited, can help achieve robustness against both stragglers and Byzantine faulty agents.  
We have presented empirical results illustrating improved communication efficacy of our asynchronous algorithm compared to the synchronous counterpart.

\paragraph{Discussion on limitations:} It is worth noting that in our results regarding convergence rate in Theorem~\ref{thm:conv-rate}, the radius $R$ of the ball bounding the estimate $x^t$ is linearly associated with $\Gamma$, or the size of the set $\W$. $\Gamma$ is arbitrary according to our definition. In our results regarding asymptotic convergence in Theorems~\ref{thm:approx}, \ref{thm:exact}, \ref{thm:async-fault-toler-cge}, and \ref{thm:approx-generalized}, the approximation bound is linearly associated with the number of all agents $n$. Therefore, such bounds might be loose when $\Gamma$ or $n$ is large. In practice, though, $R$ and the approximation bounds are also proportional to $\epsilon$ which can be small, and $\Gamma$ can also be small. 

Another worth-noting issue is that most of our results in this paper are proved under strongly-convex assumptions. One may argue that such kind of assumptions are too strong to be realistic. However, previous research has pointed out that although not a global property, cost functions of many machine learning problems are strongly-convex in the neighborhood of local minimizers \cite{bottou2018optimization}. Also, there is other research showing that the results on non-convex cost functions can be derived from those on strongly-convex cost functions \cite{allen2016optimal}, and therefore our results can be applied to a broader range of real-world problems. 
Our empirical results showing effectiveness of our algorithm also concur with the above analyses.

%% file: appendix-lemma.tex



\section{Appendix: Proof of Lemma~\ref{lemma:bound}}
\label{appdx:proof-lemma-bound}

\noindent \fbox{\begin{minipage}{\linewidth}
\textbf{Lemma~\ref{lemma:bound}.}
\textit{ Consider the general iterative update rule (\ref{eqn:update-general}). Let $\eta_t$ satisfy $\sum_{t=0}^\infty\eta_t=\infty~\textrm{ and }~\sum_{t=0}^\infty\eta_t^2<\infty$.
    For any given gradient aggregation rule $\mathsf{GradAgg}\left(\nabla Q_1(x^t), ..., \nabla Q_n(x^t)\right)$, if there exists $\M<\infty$ such that
    \begin{equation}
        \norm{\mathsf{GradAgg}\left(\nabla Q_1(x^t), ..., \nabla Q_n(x^t)\right)}\leq\M
        \tag{\ref{eqn:bounded-aggregator-norm}}
    \end{equation} for all $t$, and there exists $\D^*\in\left[0,\max_{x\in\W}\norm{x-x^*}\right)$ and $\xi>0$ such that when $\norm{x^t-x^*}\geq\D^*$,
    \begin{equation}
        \phi_t\triangleq\iprod{x^t-x^*}{\mathsf{GradAgg}\left(\nabla Q_1(x^t), ..., \nabla Q_n(x^t)\right)}\geq\xi,~
        \tag{\ref{eqn:phi-def}}
    \end{equation}
    we have $\lim_{t\rightarrow\infty}\norm{x^t-x^*}\leq\D^*$.} 
\end{minipage}}

Consider the iterative process \eqref{eqn:update-general}
\begin{equation*}
    x^{t+1}=\left[x^t-\eta_t\,\mathsf{GradAgg}\left(\nabla Q_1(x^t), ..., \nabla Q_n(x^t)\right)\right]_\W,~\forall t\geq0,
\end{equation*}
where $\W$ is a compact set. Assume that $x^*\in\W$. From now on, we use $\mathsf{GradAgg}[t]$ as a short hand for the output of the gradient aggregation rule at iteration $t$, i.e. 
\begin{equation}
    \label{eqn:def-gradagg-t}
    \mathsf{GradAgg}[t]\triangleq\mathsf{GradAgg}\left(\nabla Q_1(x^t), ..., \nabla Q_n(x^t)\right).
\end{equation}

The proof of this lemma uses the following sufficient criterion for the convergence of non-negative sequences:
\begin{lemma}[Ref. \cite{bottou1998online}]
    \label{lemma:converge}
    Consider a sequence of real values $\{u_t\}$, $t=0,1,\dots$. If $u_t\geq0$, $\forall t$, 
    \begin{align*}
        &\sum_{t=0}^\infty(u_{t+1}-u_t)_+<\infty ~\textrm{ implies }~\left\{\begin{array}{l}
            u_t\xrightarrow[t\rightarrow\infty]{}u_\infty<\infty, \\
            \sum_{t=0}^\infty(u_{t+1}-u_t)_->-\infty,
        \end{array}\right.
    \end{align*}
    where the operators $(\cdot)_+$ and $(\cdot)_-$ are defined as follows for a real value scalar $x$:
    \begin{align*}
        (x)_+=\left\{\begin{array}{ll}
            x, & x>0, \\
            0, & \textrm{otherwise},
        \end{array}\right.~\textrm{ and }~
        (x)_-=\left\{\begin{array}{ll}
            0, & x>0, \\
            x, & \textrm{otherwise}.
        \end{array}\right.
    \end{align*}
\end{lemma}
\subsection{Proof of Lemma~\ref{lemma:bound}}
Let $e_t$ denote $\norm{x^t-x^*}$. Define a scalar function $\psi$,
\begin{equation}
    \psi(y)=\left\{\begin{array}{ll}
        0, & y<\left(\D^*\right)^2, \\
        \left(y-\left(\D^*\right)^2\right)^2, & \textrm{otherwise}.
    \end{array}\right.
    \label{eqn:psi_def}
\end{equation}
Let $\psi'(y)$ denote the derivative of $\psi$ at $y$. Then (cf. \cite{bottou1998online})
\begin{equation}
    \label{eqn:psi_bnd}
    \psi(z)-\psi(y)\leq(z-y)\psi'(y)+(z-y)^2,~\forall y,z\in\mathbb{R}_{\geq0}.
\end{equation}
Note,
\begin{equation}
    \label{eqn:psi_prime}
    \psi'(y)=\max\left\{0,2\left(y-\left(\D^*\right)^2\right)\right\}.
\end{equation}

Now, define
\begin{equation}
    \label{eqn:ht_def}
    h_t\triangleq\psi(e_t^2).
\end{equation}
From \eqref{eqn:psi_bnd} and \eqref{eqn:ht_def},
\begin{equation}
    h_{t+1}-h_t=\psi\left(e_{t+1}^2\right)-\psi\left(e_t^2\right)\leq\left(e_{t+1}^2-e_t^2\right)\psi'\left(e_t^2\right)+\left(e_{t+1}^2-e_t^2\right)^2,~\forall t\in\mathbb{Z}_{\geq0}.
\end{equation}
From now on, we use $\psi_t'$ as the shorthand for $\psi'\left(e_t^2\right)$, i.e.,
\begin{equation}
    \label{eqn:def-psi-t}
    \psi_t'\triangleq\psi'\left(e_t^2\right).
\end{equation}
From above, for all $t\geq0$,
\begin{equation}
    \label{eqn:ht_1}
    h_{t+1}-h_t\leq\left(e_{t+1}^2-e_t^2\right)\psi'_t+\left(e_{t+1}^2-e_t^2\right)^2.
\end{equation}
Recall the iterative process \eqref{eqn:update}. Using the non-expansion property of Euclidean projection onto a closed convex set\footnote{$\norm{x-x^*}\geq\norm{[x]_\W-x^*},~\forall w\in\mathbb{R}^d$.}, 
\begin{equation}
    \norm{x^{t+1}-x^*}\leq\norm{x^t-\eta_t\mathsf{GradAgg}[t]-x^*}.
\end{equation}
Taking square on both sides, and recalling that $e_t\triangleq\norm{x^t-x^*}$,
\begin{equation*}
    e_{t+1}^2\leq e_t^2-2\eta_t\iprod{x_t-x^*}{\mathsf{GradAgg}[t]}+\eta_t^2\norm{\mathsf{GradAgg}[t]}^2.
\end{equation*}
Recall from \eqref{eqn:phi-def} that $\iprod{x_t-x^*}{\mathsf{GradAgg}[t]}=\phi_t$, therefore,
\begin{equation}
    \label{eqn:proj_bound}
    e_{t+1}^2\leq e_t^2-2\eta_t\phi_t+\eta_t^2\norm{\mathsf{GradAgg}[t]}^2,~\forall t\geq0.
\end{equation}

As $\psi'_t\geq0,~\forall t\in\mathbb{Z}_{\geq0}$, combining \eqref{eqn:ht_1} and \eqref{eqn:proj_bound},
\begin{equation}
    \label{eqn:ht_2}
    h_{t+1}-h_t\leq\left(-2\eta_t\phi_t+\eta_t^2\norm{\mathsf{GradAgg}[t]}^2\right)\psi'_t+\left(e_{t+1}^2-e_t^2\right)^2,~\forall t\geq0.
\end{equation}
Note that 
\begin{equation}
    \mnorm{e_{t+1}^2-e_t^2}=(e_{t+1}+e_t)\mnorm{e_{t+1}-e_t}.
\end{equation}
As $\W$ is assumed compact, there exists
\begin{equation}
    \Gamma=\max_{x\in\W}\norm{x-x^*}\leq\infty.
\end{equation}
Let $\Gamma>0$, since otherwise $\W=\{x^*\}$ only contains one point, and the problem becomes trivial. As $x^t\in\W$, $\forall t\geq0$,
\begin{equation}
    \label{eqn:e_t_bound}
    e_t\leq\Gamma,
\end{equation}
which implies
\begin{equation}
    e_{t+1}+e_t\leq2\Gamma.
\end{equation}
Therefore,
\begin{equation}
    \mnorm{e_{t+1}^2-e_t^2}\leq2\Gamma\mnorm{e_{t+1}-e_t},~\forall t\geq0.
    \label{eqn:e2t-bound-1}
\end{equation}
By triangle inequality,
\begin{equation}
    \mnorm{e_{t+1}-e_t}=\mnorm{\norm{x^{t+1}-x^*}-\norm{x^t-x^*}}\leq\norm{x^{t+1}-x^t}.
    \label{eqn:e2t-bound-2}
\end{equation}
From \eqref{eqn:update} and the non-expansion property of Euclidean projection onto a closed convex set,
\begin{equation}
    \norm{x^{t+1}-x^t}=\norm{\left[x^t-\eta_t\mathsf{GradAgg}[t]\right]_\W-x^t}\leq\eta_t\norm{\mathsf{GradAgg}[t]}.
    \label{eqn:e2t-bound-3}
\end{equation}
So from \eqref{eqn:e2t-bound-1}, \eqref{eqn:e2t-bound-2}, and \eqref{eqn:e2t-bound-3},
\begin{align}
    &\mnorm{e_{t+1}^2-e_t^2}\leq2\eta_t\Gamma\norm{\mathsf{GradAgg}[t]}, \nonumber \\
    \Longrightarrow&\left(e_{t+1}^2-e_t^2\right)^2\leq4\eta_t^2\Gamma^2\norm{\mathsf{GradAgg}[t]}^2.
\end{align}
Substituting above in \eqref{eqn:ht_2},
\begin{align}
    \label{eqn:ht_3}
    h_{t+1}-h_t\leq&\left(-2\eta_t\phi_t+\eta_t^2\norm{\mathsf{GradAgg}[t]}^2\right)\psi'_t+4\eta_t^2\Gamma^2\norm{\mathsf{GradAgg}[t]}^2, \nonumber \\
    =&-2\eta_t\phi_t\psi'_t+(\psi'_t+4\Gamma^2)\eta_t^2\norm{\mathsf{GradAgg}[t]}^2,~\forall t\geq0.
\end{align}

Recall that the statement of Lemma~\ref{lemma:bound} assumes that $\D^*\in[0,\max_{x\in\W}\norm{x-x^*})$, which indicates $\D^*<\Gamma$. Using \eqref{eqn:psi_prime} and \eqref{eqn:e_t_bound}, we have
\begin{equation}
    0\leq\psi'_t\leq2\left(e_t^2-\left(\D^*\right)^2\right)\leq2\left(\Gamma^2-\left(\D^*\right)^2\right)\leq2\Gamma^2.
    \label{eqn:psi_prime_t_bnd}
\end{equation}
Recall from \eqref{eqn:bounded-aggregator-norm} that  $\norm{\mathsf{GradAgg}[t]}\leq\M<\infty$ for all $t$. Substituting \eqref{eqn:psi_prime_t_bnd} in \eqref{eqn:ht_3},
\begin{align}
    \label{eqn:ht_4}
    h_{t+1}-h_t\leq&-2\eta_t\phi_t\psi'_t+(2\Gamma^2+4\Gamma^2)\eta_t^2\M^2=-2\eta_t\phi_t\psi'_t+6\Gamma^2\eta_t^2\M^2,~\forall t\geq0.
\end{align}

Now we use Lemma~\ref{lemma:converge} to show that $h_\infty=0$ as follows. For each iteration $t$, consider the following two cases:
\begin{description}
    \item[Case 1)] Suppose $e_t<\D^*$. In this case, $\psi'_t=0$. By Cauchy-Schwartz inequality,
        \begin{equation}
            \mnorm{\phi_t}=\mnorm{\iprod{x^t-x^*}{\mathsf{GradAgg}[t]}}\leq e_t\norm{\mathsf{GradAgg}[t]}.
        \end{equation}
        By \eqref{eqn:bounded-aggregator-norm} and \eqref{eqn:e_t_bound}, this implies that
        \begin{equation}
            \mnorm{\phi_t}\leq\Gamma\M<\infty.
        \end{equation}
        Thus,
        \begin{equation}
            \label{eqn:phitpsit_1}
            \phi_t\psi'_t=0.
        \end{equation}
    \item[Case 2)] Suppose $e_t\geq\D^*$. Therefore, there exists $\delta\geq0$, $e_t=\D^*+\delta$. From \eqref{eqn:psi_prime}, we obtain that
    \begin{equation}
        \psi_t'=2\left(\left(\D^*+\delta\right)^2-\left(\D^*\right)^2\right)=2\delta\left(2\D^*+\delta\right).
    \end{equation}
    The statement of Lemma~\ref{lemma:bound} assumes that $\phi_t\geq\xi>0$ when $e_t\geq\D^*$, thus, 
        \begin{equation}
            \label{eqn:phitpsit_2}
            \phi_t\psi'_t\geq2\delta\xi\left(2\D^*+\delta\right)>0.
        \end{equation}
\end{description}
From \eqref{eqn:phitpsit_1} and \eqref{eqn:phitpsit_2}, for both cases,
\begin{equation}
    \phi_t\psi'_t\geq0, ~\forall t\geq0.
    \label{eqn:phi_psi_bound}
\end{equation}
Combining this with \eqref{eqn:ht_4},
\begin{equation}
    h_{t+1}-h_t\leq6\Gamma^2\eta_t^2\M^2.
\end{equation}
From above we have
\begin{equation}
    \left(h_{t+1}-h_t\right)_+\leq6\Gamma^2\eta_t^2\M^2.
\end{equation}
Since $\sum_{t=0}^\infty\eta_t^2<\infty$, $\Gamma,\M<\infty$,
\begin{equation}
    \sum_{t=0}^\infty\left(h_{t+1}-h_t\right)_+\leq6\Gamma^2\M^2\sum_{t=0}^\infty\eta_t^2<\infty.
\end{equation}
Then Lemma~\ref{lemma:converge} implies that by the definition of $h_t$, we have $h_t\geq0,~\forall t$, 
\begin{align}
    h_t\xrightarrow[t\rightarrow\infty]{}h_\infty<\infty,~\textrm{and} \label{eqn:upper_bound_h_infty}\\
    \sum_{t=0}^\infty\left(h_{t+1}-h_t\right)_->-\infty.
\end{align}
Note that $h_\infty-h_0=\sum_{t=0}^\infty(h_{t+1}-h_t)$. Thus, from \eqref{eqn:ht_4} we have 
\begin{equation}
    h_\infty-h_0\leq-2\sum_{t=0}^\infty\eta_t\phi_t\psi_t'+6\Gamma^2\M^2\sum_{t=0}^\infty\eta_t^2.
\end{equation}
By \eqref{eqn:ht_def} the definition of $h_t$, $h_t\geq0$ for all $t$. Therefore, from above we obtain
\begin{equation}
    2\sum_{t=0}^\infty\eta_t\phi_t\psi_t'\leq h_0-h_\infty+6\Gamma^2\M^2\sum_{t=0}^\infty\eta_t^2.
    \label{eqn:bound_2_sum}
\end{equation}
By assumption, $\sum_{t=0}^\infty\eta_t^2<\infty$. Substituting from \eqref{eqn:e_t_bound} that $e_t<\infty$ in \eqref{eqn:ht_def}, we obtain that 
\begin{equation}
    h_0=\psi\left(e_0^2\right)<\infty.
\end{equation} 
Therefore, \eqref{eqn:bound_2_sum} implies that
\begin{equation}
    2\sum_{t=0}^\infty\eta_t\phi_t\psi_t'\leq h_0+6\Gamma^2\M^2\sum_{t=0}^\infty\eta_t^2<\infty.
\end{equation}
Or simply,
\begin{equation}
    \sum_{t=0}^\infty\eta_t\phi_t\psi_t'<\infty.
    \label{eqn:upper_bound_etatphitpsit}
\end{equation}

Finally, we reason below by contradiction that $h_\infty=0$. Note that for any $\zeta>0$, there exists a unique positive value $\beta$ such that $\zeta=2\beta\left(2\D^*+\sqrt{\beta}\right)^2$. Suppose that $h_\infty=2\beta\left(2\D^*+\sqrt{\beta}\right)^2$ for some positive value $\beta$. As the sequence $\{h_t\}_{t=0}^\infty$ converges to $h_\infty$ (see \eqref{eqn:upper_bound_h_infty}), there exists some finite $\tau\in\Z_{\geq0}$ such that for all $t\geq\tau$, 
\begin{align}
    &\mnorm{h_t-h_\infty}\leq\beta\left(2\D^*+\sqrt{\beta}\right)^2 \\
    \Longrightarrow & h_t\geq h_\infty-\beta\left(2\D^*+\sqrt{\beta}\right)^2.
\end{align}
As $h_\infty=2\beta\left(2\D^*+\sqrt{\beta}\right)^2$, the above implies that
\begin{equation}
    h_t\geq \beta\left(2\D^*+\sqrt{\beta}\right)^2, \forall t\geq\tau.
    \label{eqn:ht_lower_bound}
\end{equation}
Therefore (cf. \eqref{eqn:psi_def} and \eqref{eqn:ht_def}), for all $t\geq\tau$,
\begin{eqnarray*}
    \left(e_t^2-\left(\D^*\right)^2\right)^2\geq\beta\left(2\D^*+\sqrt{\beta}\right)^2, \textrm{ or} \\
    \mnorm{e_t^2-\left(\D^*\right)^2}\geq\sqrt{\beta}\left(2\D^*+\sqrt{\beta}\right).
\end{eqnarray*}
Thus, for each $t\geq\tau$, either
\begin{equation}
    e^2_t\geq\left(\D^*\right)^2+\sqrt{\beta}\left(2\D^*+\sqrt{\beta}\right)=\left(\D^*+\sqrt{\beta}\right)^2,
    \label{eqn:et_case_1}
\end{equation}
or
\begin{equation}
    e^2_t\leq\left(\D^*\right)^2-\sqrt{\beta}\left(2\D^*+\sqrt{\beta}\right)<\left(\D^*\right)^2.
    \label{eqn:et_case_2}
\end{equation}
If the latter, i.e., \eqref{eqn:et_case_2} holds true for some $t'\geq\tau$, 
\begin{equation}
    h_{t'}=\psi\left(e_{t'}^2\right)=0,
\end{equation}
which contradicts \eqref{eqn:ht_lower_bound}. Therefore, \eqref{eqn:ht_lower_bound} implies \eqref{eqn:et_case_1}.

From above we obtain that if $h_\infty=2\beta\left(2\D^*+\sqrt{\beta}\right)^2$, there exists $\tau<\infty$ such that for all $t\geq\tau$, 
\begin{equation}
    e_t\geq\D^*+\sqrt{\beta}.
\end{equation}
Thus, from \eqref{eqn:phitpsit_2}, with $\delta=\sqrt{\beta}$, we obtain that 
\begin{equation}
    \phi_t\psi_t'\geq2\xi\sqrt{\beta}\left(2\D^*+\sqrt{\beta}\right), \forall t\geq\tau.
\end{equation}
Therefore,
\begin{equation}
    \sum_{t=\tau}^\infty\eta_t\phi_t\psi_t'\geq2\xi\sqrt{\beta}\left(2\D^*+\sqrt{\beta}\right)\sum_{t=\tau}^\infty\eta_t=\infty.
\end{equation}
This is a contradiction to \eqref{eqn:upper_bound_etatphitpsit}. Therefore, $h_\infty=0$, and by \eqref{eqn:ht_def}, the definition of $h_t$, 
\begin{equation}
    h_\infty=\lim_{t\rightarrow\infty}\psi\left(e_t^2\right)=0.
\end{equation}
Hence, by \eqref{eqn:psi_def}, the definition of $\psi(\cdot)$, 
\begin{equation}
    \lim_{t\rightarrow\infty}\norm{x^t-x^*}\leq\D^*.
\end{equation}

%% file: appendix-main.tex
\section{Appendix: Proofs of convergence}
\label{appdx:proof-main}

\subsection{Proof for Theorem~\ref{thm:approx}}
\label{appdx-sub:proof-approx}

\noindent \fbox{\begin{minipage}{\linewidth}
\textbf{Theorem~\ref{thm:approx}. }
    \textit{ Suppose that Assumptions~\ref{assum:lipschitz} and \ref{assum:strongly-convex} hold true, and the cost functions of all agents satisfies $(r,\epsilon)$-redundancy. 
     Assume that $\eta_t$ in (\ref{eqn:update}) satisfies $\sum_{t=0}^\infty\eta_t=\infty$ and $\sum_{t=0}^\infty\eta_t^2<\infty$.
     Let $\alpha$ and $D$ be defined as follows: 
     \[\alpha\triangleq 1-\dfrac{r}{n}\cdot\dfrac{\mu}{\gamma}>0~\textrm{ and }~\D\triangleq\dfrac{2r\mu}{\alpha\gamma}\epsilon.\]
     Then, for the proposed Algorithm~\ref{alg}, 
    \begin{equation*}
        \lim_{t\rightarrow\infty}\norm{x^t-x^*}\leq\D, 
    \end{equation*}
    where $x^*=\arg\min_x\sum_{i\in[n]}Q_i(x)$.}
\end{minipage}}

\textbf{First}, we need to show that $\norm{\sum_{j\in S^t}\nabla Q_j(x^t)}$ is bounded for all $t$. By Assumption~\ref{assum:lipschitz}, for all $j\in[n]$, 
\begin{equation}
    \label{eqn:thm2-lip}
    \norm{\nabla Q_j(x)-\nabla Q_j(x^*)}\leq\mu\norm{x-x^*}.
\end{equation}

Let $x_S=\arg\min_x\sum_{j\in S}Q_j(x)$ be the minimum point of the aggregated cost functions of a set $S$ of $n-r$ agents, i.e., $\mnorm{S}=n-f$. Note that $\sum_{j\in S}\nabla Q_j(x_S)=0$.
By triangle inequality,
\begin{equation}
    \label{eqn:thm2-triangle}
    \norm{\sum_{j\in S}\nabla Q_j(x_S)-\sum_{j\in S}\nabla Q_j(x^*)}\leq\sum_{j\in S}\norm{\nabla Q_j(x_S)-\nabla Q_j(x^*)}.
\end{equation}
Combining~\eqref{eqn:thm2-lip} through \eqref{eqn:thm2-triangle}, 
\begin{equation}
    \norm{\sum_{j\in S}\nabla Q_j(x^*)}\leq\mnorm{S}\mu\norm{x_S-x^*}.
\end{equation}
By Definition~\ref{def:approx_redundancy} of $(r,\epsilon)$-redundancy, $\norm{x_S-x^*}\leq\epsilon$.
Therefore,
\begin{equation}
    \norm{\sum_{j\in S}\nabla Q_j(x^*)}\leq\mu\epsilon (n-r).
\end{equation}

Now, consider an arbitrary agent $i\in[n]\backslash S$. Let $T=S\cup\{i\}$. Using a similar argument as above, we obtain 
\begin{equation}
    \norm{\sum_{j\in T}\nabla Q_j(x^*)}\leq\mu\epsilon (n-r+1).
\end{equation}
Therefore, 
\begin{align}
    \norm{\nabla Q_i(x^*)}=&\norm{\sum_{j\in T}\nabla Q_j(x^*)-\sum_{j\in S}\nabla Q_j(x^*)} \leq\norm{\sum_{j\in T}\nabla Q_j(x^*)}+\norm{\sum_{j\in S}\nabla Q_j(x^*)} \nonumber \\
        \leq&(n-r)\mu\epsilon+(n-r+1)\mu\epsilon=(2n-2r+1)\mu\epsilon. \label{eqn:apprx_gradient_bnd}
\end{align}
Note that the inequality \eqref{eqn:apprx_gradient_bnd} can be applied to any $i\in[n]$ with a suitable choice of $S$ above.\\

On the other hand, by Assumption~\ref{assum:lipschitz}, for any $x\in\mathbb{R}^d$,
\begin{equation}
    \norm{\nabla Q_i(x)-\nabla Q_i(x^*)}\leq\mu\norm{x-x^*}.
\end{equation}
By triangle inequality,
\begin{equation}
    \norm{\nabla Q_i(x)}\leq\norm{\nabla Q_i(x^*)}+\mu\norm{x-x^*}.
\end{equation}
Combining above and \eqref{eqn:apprx_gradient_bnd},
\begin{equation}
    \label{eqn:apprx_gradient_bnd_2}
    \norm{\nabla Q_i(x)}\leq(2n-2r+1)\mu\epsilon+\mu\norm{x-x^*}\leq2n\mu\epsilon+\mu\norm{x-x^*}.
\end{equation}
Recall that $x^t\in\W$ for all $t$, where $\W$ is a compact set. There exists a $\Gamma=\max_{x\in\W}\norm{x-x^*}<\infty$, such that $\norm{x^t-x^*}\leq\Gamma$ for all $t$. 
Therefore, for all $t$, 
\begin{align}
    \label{eqn:apprx_phi_t_bnd}
    \norm{\sum_{j\in S^t}\nabla Q_j(x^t)}&=\norm{\sum_{j\in S^t}\nabla Q_j(x^t)}\leq\sum_{j\in S^t}\norm{\nabla Q_j(x^t)} \nonumber \\
        &\leq\mnorm{S^t}\cdot\left(2n\mu\epsilon+\mu\norm{x^t-x^*}\right)\leq(n-r)\left(2n\mu\epsilon+\mu\Gamma\right)<\infty.
\end{align}

\textbf{Second}, consider the following term $\displaystyle\phi_t=\iprod{x^t-x^*}{\sum_{j\in S^t}\nabla Q_j(x^t)}$.
We have
\begin{align}
    \label{eqn:apprx_phi_t_1}
    \phi_t=&\iprod{x^t-x^*}{\sum_{j\in S^t}\nabla Q_j(x^t)} \nonumber \\
        =&\iprod{x^t-x^*}{\sum_{j\in S^t}\nabla Q_j(x^t)+\sum_{k\in [n]\backslash S^t}\nabla Q_k(x^t)-\sum_{k\in [n]\backslash S^t}\nabla Q_k(x^t)} \nonumber \\
        =&\iprod{x^t-x^*}{\sum_{j\in [n]}\nabla Q_j(x^t)}-\iprod{x^t-x^*}{\sum_{k\in [n]\backslash S^t}\nabla Q_k(x^t)}.
\end{align}

For the first term in \eqref{eqn:apprx_phi_t_1}, recall that $Q_S(x)=\frac{1}{\mnorm{S}}\sum_{j\in S}Q_j(x)$ for any set $S$ that satisfies $\mnorm{S}\geq n-r$. With Assumption~\ref{assum:strongly-convex}, 
\begin{align}
    \label{eqn:apprx_phi_t_1_1}
    &\iprod{x^t-x^*}{\sum_{j\in [n]}\nabla Q_j(x^t)}=\iprod{x^t-x^*}{\sum_{j\in [n]}\nabla Q_j(x^t)-\sum_{j\in [n]}\nabla Q_j(x^*)} \nonumber \\
        =&n\cdot\iprod{x^t-x^*}{\nabla Q_{[n]}(x^t)-\nabla Q_{[n]}(x^*)}\geq n\gamma\norm{x^t-x^*}^2.
\end{align}

For the second term in \eqref{eqn:apprx_phi_t_1}, by Cauchy-Schwartz inequality,
\begin{align}
    \label{eqn:apprx_phi_t_1_2}
    \iprod{x^t-x^*}{\sum_{k\in [n]\backslash S^t}\nabla Q_k(x^t)}&=\sum_{k\in [n]\backslash S^t}\iprod{x^t-x^*}{\nabla Q_k(x^t)}\nonumber \\
        &\leq\sum_{k\in [n]\backslash S^t}\norm{x^t-x^*}\cdot\norm{\nabla Q_k(x^t)}.
\end{align}

Combining \eqref{eqn:apprx_phi_t_1}, \eqref{eqn:apprx_phi_t_1_1}, and \eqref{eqn:apprx_phi_t_1_2},
\begin{equation}
    \label{eqn:apprx_phi_t_2}
    \phi_t\geq n\gamma\norm{x^t-x^*}^2-\sum_{k\in[n]\backslash S^t}\norm{x^t-x^*}\norm{\nabla Q_k(x^t)}.
\end{equation}
Substituting from \eqref{eqn:apprx_gradient_bnd_2} in above, note that $\mnorm{[n]\backslash S^t}=r$,
\begin{align}
    \label{eqn:apprx_phi_t_3}
    \phi_t&\geq n\gamma\norm{x^t-x^*}^2-\sum_{k\in[n]\backslash S^t}\norm{x^t-x^*}(2n\mu\epsilon+\mu\norm{x^t-x^*}) \nonumber \\
        &=n\gamma\norm{x^t-x^*}^2-r\norm{x^t-x^*}(2n\mu\epsilon+\mu\norm{x^t-x^*}) \nonumber \\
        &=n\gamma\left(1-\dfrac{r}{n}\cdot\dfrac{\mu}{\gamma}\right)\norm{x^t-x^*}\left(\norm{x^t-x^*}-\dfrac{2r\mu\epsilon}{\gamma\left(1-\dfrac{r}{n}\cdot\dfrac{\mu}{\gamma}\right)}\right).
\end{align}
Recall that we assume $\displaystyle\alpha=1-\dfrac{r}{n}\cdot\dfrac{\mu}{\gamma}>0$.
We have
\begin{equation}
    \phi_t\geq\alpha n\gamma\norm{x^t-x^*}\left(\norm{x^t-x^*}-\dfrac{2r\mu}{\alpha\gamma}\epsilon\right).
    \label{eqn:phi_t_final}
\end{equation}
Let $\D\triangleq\dfrac{2r\mu}{\alpha\gamma}\epsilon$. \eqref{eqn:phi_t_final} implies for an arbitrary $\delta>0$, 
\begin{equation}
    \phi_t\geq\alpha n\gamma\delta\left(D+\delta\right)>0~\textrm{ when }~\norm{x^t-x^*}\geq\D+\delta. 
\end{equation}
Therefore, by Lemma~\ref{lemma:bound}, 
\begin{equation}
    \lim_{t\rightarrow\infty}\norm{x^t-x^*}\leq\D.
\end{equation}

\subsection{Proof for Lemma~\ref{lemma:minimum}}
\label{appdx-sub:proof-minimum-lemma}

\noindent \fbox{\begin{minipage}{\linewidth}
\textbf{Lemma~\ref{lemma:minimum}.} 
\textit{Under Assumptions~\ref{assum:lipschitz} and \ref{assum:strongly-convex}, $\nabla Q_j(x^*)=0$ for all $j\in[n]$, if the cost functions of all agents satisfies $r$-redundancy property.}
\end{minipage}}

By Definition~\ref{def:redundancy}, the minimum point $x^*=\arg\min_x\sum_{i\in[n]}Q_i(x)$ also satisfies $x^*=\arg\min_x\sum_{i\in S}Q_i(x)$, for any $\mnorm{S}\geq n-r$. 

Consider any set $S$ with $\mnorm{S}=n-r$ and any agent $j\not\in S$. Let $T=S\cup\{j\}$.
$S$ and $T$ satisfy (\ref{eqn:redundancy}), so 
\begin{equation}
    \sum_{i\in S}\nabla Q_i(x^*)=\sum_{i\in T}\nabla Q_i(x^*) = 0.
\end{equation}
Note that
\begin{equation}
    \sum_{i\in T}\nabla Q_i(x) = \sum_{i\in S}\nabla Q_i(x) + \nabla Q_j(x)
\end{equation}
Combine the two equations together, we have
\begin{equation}
    \nabla Q_j(x) = 0.
\end{equation}
Notice that the choice of $S$, $T$ and $j$ are arbitrary. Therefore, $\nabla Q_i(x^*)=0$, $\forall i\in[n]$.

\subsection{Proof for Theorem~\ref{thm:exact}}
\label{appdx-sub:proof-exact}

\noindent \fbox{\begin{minipage}{\linewidth}
\textbf{Theorem~\ref{thm:exact}.}
    \textit{Suppose Assumptions~\ref{assum:lipschitz} and \ref{assum:strongly-convex} hold true, and the cost functions of all agents satisfies $r$-redundancy. Assume that $\eta_t$ in (\ref{eqn:update}) satisfies $\sum_{t=0}^\infty\eta_t=\infty$ and $\sum_{t=0}^\infty\eta_t^2<\infty$, Algorithm~\ref{alg} has 
    \begin{equation*}
        \lim_{t\rightarrow\infty}x^t=x^*, 
    \end{equation*}
    where $x^*=\arg\min_x\sum_{i\in[n]}Q_i(x)$.}
\end{minipage}}

\textbf{First}, we show that $\norm{\sum_{j\in S^t}\nabla Q_j(x^t)}$ is bounded for all $t$. By Assumption~\ref{assum:lipschitz}, for all $i$, $\norm{\nabla Q_i(x^t)-\nabla Q_i(x^*)}\leq\mu\norm{x^t-x^*}$.
Combining it with Lemma~\ref{lemma:minimum}, we have 
  \begin{equation}
    \norm{\nabla Q_i(x^t)}\leq\mu\norm{x^t-x^*}.
\end{equation}
By triangle inequality,
\begin{align}
    \norm{\sum_{j\in S^t}\nabla Q_j(x^t)}\leq\sum_{j\in S^t}\norm{\nabla Q_j(x^t)}\leq\mnorm{S^t}\cdot\mu\norm{x^t-x^*}.
\end{align}
Recall that $x^t\in\W$ for all $t$, where $\W$ is a compact set. There exists a $\Gamma=\max_{x\in\W}\norm{x-x^*}<\infty$, such that $\norm{x^t-x^*}\leq\Gamma$ for all $t$. Also note that $|S^t|=n-r$. Therefore, we have
\begin{align}
    \norm{\sum_{j\in S^t}\nabla Q_j(x^t)}\leq(n-r)\mu\Gamma,~\forall t.
\end{align}
Therefore, $\norm{\sum_{j\in S^t}\nabla Q_j(x^t)}$ is bounded for all $t$.

\textbf{Second}, consider the following term 
\begin{equation}
    \phi_t=\iprod{x^t-x^*}{\sum_{j\in S^t}\nabla Q_j(x^t)}.
\end{equation}
Recall Lemma~\ref{lemma:minimum}, $\nabla Q_j(x^*)=0$ for all $j\in[n]$. With Assumption~\ref{assum:strongly-convex}, 
\begin{align}
    \phi_t&=\iprod{x^t-x^*}{\sum_{j\in S^t}\nabla Q_j(x^t)}~=~\iprod{x^t-x^*}{\sum_{j\in S^t}\nabla Q_j(x^t)-\sum_{j\in S^t}\nabla Q_j(x^*)} \nonumber \\
        &=\mnorm{S^t}\cdot\iprod{x^t-x^*}{\nabla Q_{S^t}(x^t)-\nabla Q_{S^t}(x^*)} \geq(n-r)\gamma\norm{x^t-x^*}^2 
\end{align}
Therefore, for an arbitrary $\delta>0$, $\phi_t\geq(n-r)\gamma\delta$ so long as $\norm{x^t-x^*}\geq\delta$.

Applying Lemma~\ref{lemma:bound} with $\M=(n-r)\mu\Gamma$, $\xi=(n-r)\gamma\delta$, and $\D^*=\delta$, we obtain that for all $\delta>0$, $\lim_{t\rightarrow\infty}\norm{x^t-x^*}\leq\delta$. By the arbitrariness of $\delta$, 
\begin{equation}
    \lim_{t\rightarrow\infty}\norm{x^t-x^*}\leq0.
\end{equation} 
Since $\norm{x^t-x^*}$ is always non-negative,
\begin{equation}
    \lim_{t\rightarrow\infty}x^t=x^*.
\end{equation}

%% file: appendix-rate.tex
\section{Appendix: Proofs for Theorem~\ref{thm:conv-rate}}

\noindent \fbox{\begin{minipage}{\linewidth}
\textbf{Theorem~\ref{thm:conv-rate}.} 
    \textit{Suppose Assumptions~\ref{assum:lipschitz} and \ref{assum:strongly-convex} hold, and the agents' cost functions satisfy $(r,\epsilon)$-redundancy. Define $\alpha=1-\cfrac{r}{n}\cdot\cfrac{\mu}{\gamma}$ and $\overline{\eta}=\cfrac{2\gamma\alpha}{\mu^2 n}$. 
    There exists a positive real-value $R \in \Theta(\epsilon)$, such that Algorithm~\ref{alg} satisfy (i) $A\in[0,1)$, and (ii) $\norm{x^t-x^*}^2\leq A^t\norm{x^0-x^*}+R$, if
    \begin{enumerate}[nosep,label=\alph*),leftmargin=*]
        \item let $A=1-(\mu n)^2\eta(\overline{\eta}-\eta)$ for some $\eta$, and $\eta_t=\eta$ for every iteration $t$ with $\eta\in(0,\overline{\eta})$.  \label{thm:conv-rate-a}
        \item let $A=1-(\mu n)^2c(\overline{\eta}-c)$ for some $c$, and $\eta_t=c/(t+1)$ for every iteration $t$ with $c\in(0,\overline{\eta})$. \label{thm:conv-rate-b}
    \end{enumerate}}
\end{minipage}}

\subsection{Proof for Theorem~\ref{thm:conv-rate}\ref{thm:conv-rate-a}}
\label{appdx:proof-rate-a}

Define $\g^t=\sum_{j\in S^t}\nabla Q_j(x^t)$. 
Recall our iterative update \eqref{eqn:update}. Using the non-expansion property of Euclidean projection onto a closed convex set\footnote{$\norm{x-x^*}\geq\norm{[x]_\W-x^*},~\forall w\in\mathbb{R}^d$.}, 
\begin{equation}
    \label{eqn:apprx-one-step}
    \norm{x^{t+1}-x^*}\leq\norm{x^t-\eta\g^t-x^*}.
\end{equation}

Following the argument in the \textbf{first} part of the proof of Theorem~\ref{thm:approx}, by Assumption~\ref{assum:lipschitz} and $(r,\epsilon)$-redundancy, 
\begin{equation}
    \norm{\nabla Q_i(x)}\leq2n\mu\epsilon+\mu\norm{x-x^*}. \tag{\ref{eqn:apprx_gradient_bnd_2}}
\end{equation}
Therefore, by triangle inequality, since $\mnorm{S^t}=n-r$, we have
\begin{equation}
    \label{eqn:update-bnd}
    \norm{\g^t}\leq\sum_{j\in S^t}\norm{\nabla Q_j(x^t)}\leq (n-r)(2n\mu\epsilon+\mu\norm{x^t-x^*}).
\end{equation}
Substituting from above in \eqref{eqn:apprx-one-step}, we obtain that
\begin{align}
    \label{eqn:rate-1}
    \norm{x^{t+1}-x^*}^2\leq&\norm{x^t-x^*}^2-2\eta_t\iprod{x^t-x^*}{\g^t}+\eta_t^2\norm{\g^t}^2 \nonumber \\
        \leq&\norm{x^t-x^*}^2-2\eta_t\iprod{x^t-x^*}{\g^t}+\eta_t^2(n-r)^2(2n\mu\epsilon+\mu\norm{x^t-x^*})^2
\end{align}
We define
\begin{equation}
    \phi_t=\iprod{x^t-x^*}{\g^t}.
\end{equation}
Following the same argument in the \textbf{second} part of the proof of Theorem~\ref{thm:approx}, by Assumption~\ref{assum:strongly-convex}, 
\begin{align}
    \phi_t&\geq n\gamma\norm{x^t-x^*}^2-r\norm{x^t-x^*}(2n\mu\epsilon+\mu\norm{x^t-x^*}) \tag{\ref{eqn:apprx_phi_t_3}}
\end{align}
Let $e_t$ denote $\norm{x^t-x^*}$, and note that $\eta_t=\eta$ for all $t$. Combining above and \eqref{eqn:rate-1}, we obtain that
\begin{align}
    \label{eqn:et-ineq-1}
    e_{t+1}^2\leq&\left[1+2\eta\left(\mu r-\gamma n\right)+\eta^2\mu^2(n-r)^2\right]e_t^2 + 4\left[\eta\mu\epsilon nr+\eta^2\mu^2\epsilon(n-r)^2n\right]e_t + 4\eta^2\mu^2\epsilon^2(n-r)^2n^2 \nonumber \\
    \leq&\left[1+2\eta\left(\mu r-\gamma n\right)+\eta^2\mu^2n^2\right]e_t^2 + 4\left(\eta\mu\epsilon n^2+\eta^2\mu^2\epsilon n^3\right)e_t + 4\eta^2\mu^2\epsilon^2n^4.
\end{align}
Let $A=1+2\eta\left(\mu r-\gamma n\right)+\eta^2\mu^2n^2$, $B=\eta\mu\epsilon nr+\eta^2\mu^2\epsilon n^3$, and $C=4\eta^2\mu^2\epsilon^2n^4$. 
Recall that $x^t\in\W$ for all $t$, where $\W$ is a compact set. There exists a $\Gamma=\max_{x\in\W}\norm{x-x^*}<\infty$, such that $\norm{x^t-x^*}\leq\Gamma$ for all $t$, or $e_t\leq\Gamma$. Thus, from \eqref{eqn:et-ineq-1} we obtain
\begin{equation}
    \label{eqn:et-ineq-2}
    e_{t+1}^2\leq Ae_t^2+4B\Gamma+C. 
\end{equation}
Let $D=4B\Gamma+C$, 
\begin{equation}
    \label{eqn:et-ineq-3}
    e_{t+1}^2\leq Ae_t^2+D.
\end{equation}
Note that $B,C,D\geq0$.

Let us first define $\alpha=1-\cfrac{r}{n}\cdot\cfrac{\mu}{\gamma}$. Consider an arbitrary set of more than $n-r$ agents $S$, and two points $x,y\in\R^d$, $x\neq y$. By Assumption~\ref{assum:lipschitz}, for every $j\in S$, 
\begin{equation}
    \norm{\nabla Q_j(x)-\nabla Q_j(y)}\leq\mu\norm{x-y}.
\end{equation}
By Cauchy-Schwartz inequality,
\begin{equation}
    \iprod{x-y}{\nabla Q_j(x)-\nabla Q_j(y)}\leq\norm{x-y}\cdot\norm{\nabla Q_j(x)-\nabla Q_j(y)}\leq\mu\norm{x-y}^2.
\end{equation}
Therefore, 
\begin{equation}
    \label{eqn:cmp-mu}
    \sum_{j\in S}\iprod{x-y}{\nabla Q_j(x)-\nabla Q_j(y)}\leq\mu\mnorm{S}\norm{x-y}^2.
\end{equation}
On the other hand, by Assumption~\ref{assum:strongly-convex}, 
\begin{equation}
    \iprod{x-y}{\nabla Q_S(x)-\nabla Q_S(y)}\geq\gamma\norm{x-y}^2.
\end{equation}
Recall that $Q_S(x)=\cfrac{1}{\mnorm{S}}\sum_{j\in S}Q_j(x)$. From above we obtain that
\begin{equation}
    \label{eqn:cmp-gamma}
    \sum_{j\in S}\iprod{x-y}{\nabla Q_j(x)-\nabla Q_j(y)}\geq\gamma\mnorm{S}\norm{x-y}^2.
\end{equation}
Compare \eqref{eqn:cmp-mu} and \eqref{eqn:cmp-gamma}, as $x$ and $y$ are arbitrary and non-identical points, suppose Assumptions~\ref{assum:lipschitz} and \ref{assum:strongly-convex} hold at the same time, we have $\gamma<\mu$. Therefore, 
\begin{equation}
    \frac{\gamma}{\mu}\cdot\alpha=\frac{\gamma}{\mu}-\frac{r}{n}\leq\frac{\gamma}{\mu}\leq1.
    \label{eqn:gamma-mu-alpha}
\end{equation}

Now, consider the value of $A$. By the definition of $\alpha$, $\mu r-\gamma n=-\gamma n\alpha$. So $A$ can be written as
\begin{equation}
    A=1-2\eta\gamma n \alpha + \eta^2\mu^2n^2=1-(\mu n)^2\eta\left(\frac{2\gamma\alpha}{\mu^2 n}-\eta\right).
\end{equation}
Let $\overline{\eta}=\cfrac{2\gamma\alpha}{\mu^2 n}$, 
from above we obtain that
\begin{equation}
    \label{eqn:def-A}
    A=1-(\mu n)^2\eta(\overline{\eta}-\eta).
\end{equation}
Note that
\begin{equation}
    \eta(\overline{\eta}-\eta)=\left(\frac{\overline{\eta}}{2}\right)^2-\left(\eta-\frac{\overline{\eta}}{2}\right)^2.
\end{equation}
Therefore, 
\begin{equation}
    A=1-(\mu n)^2\left[\left(\frac{\overline{\eta}}{2}\right)^2-\left(\eta-\frac{\overline{\eta}}{2}\right)^2\right]=(\mu n)^2\left(\eta-\frac{\overline{\eta}}{2}\right)^2+1-(\mu n)^2\left(\frac{\overline{\eta}}{2}\right)^2.
\end{equation}
Since $\eta\in(0,\overline{\eta})$, the minimum value of $A$ can be obtained when $\eta=\overline{\eta}/2$, 
\begin{equation}
    \min_{\eta}A=1-(\mu n)^2\left(\frac{\overline{\eta}}{2}\right)^2.
\end{equation}
On the other hand, since $\eta\in(0,\overline{\eta})$, from \eqref{eqn:def-A}, $A<1$. Thus,
\begin{equation}
    1-(\mu n)^2\left(\frac{\overline{\eta}}{2}\right)^2\leq A<1.
\end{equation}
Substituting $\overline{\eta}=\cfrac{2\gamma\alpha}{\mu^2n}$ in above implies that $A\in\left[1-\left(\cfrac{\gamma\alpha}{\mu}\right)^2,1\right)$. Since $\cfrac{\gamma\alpha}{\mu}\leq1$ (cf. \eqref{eqn:gamma-mu-alpha}), $A\in[0,1)$.

Knowing that $A<1$, from \eqref{eqn:et-ineq-3} we obtain by induction that
\begin{align}
    e_{t+1}^2\leq& Ae_t^2+D \nonumber \\
        \leq& A(Ae_{t-1}^2+D)+D = A^2e_{t-1}^2+AD+D \nonumber \\
        \leq&... \nonumber \\ 
        \leq& A^{t+1}e_0^2+(A^t+...+A+1)D = A^{t+1}e_0^2+\frac{1-A^t}{1-A}D.
\end{align}
Since $A<1$, $A^t<1$ for all $t$. Thus,
\begin{equation}
    e_{t+1}^2\leq A^{t+1}e_0^2+\frac{D}{1-A}.
\end{equation}
Let $R=\frac{D}{1-A}$, we have
\begin{align}
    \norm{x^t-x^*}^2\leq A^{t}\norm{x^0-x^*}^2+R.
\end{align}

Note that since $B=\Theta(\epsilon)$, $C=\Theta(\epsilon^2)$, we have $D=\Theta(\epsilon)$ and therefore, $R=\Theta(\epsilon)$.

\subsection{Proof for Theorem~\ref{thm:conv-rate}\ref{thm:conv-rate-b}}
\label{appdx:proof-rate-b}


Define $\g^t=\sum_{j\in S^t}\nabla Q_j(x^t)$. 
Recall our iterative update \eqref{eqn:update}. Using the non-expansion property of Euclidean projection onto a closed convex set\footnote{$\norm{x-x^*}\geq\norm{[x]_\W-x^*},~\forall w\in\mathbb{R}^d$.}, 
\begin{equation}
    \label{eqn:apprx-one-step-dim}
    \norm{x^{t+1}-x^*}\leq\norm{x^t-\eta\g^t-x^*}.
\end{equation}

Following the argument in the \textbf{first} part of the proof of Theorem~\ref{thm:approx}, by Assumption~\ref{assum:lipschitz} and $(r,\epsilon)$-redundancy, 
\begin{equation}
    \norm{\nabla Q_i(x)}\leq2n\mu\epsilon+\mu\norm{x-x^*}. \tag{\ref{eqn:apprx_gradient_bnd_2}}
\end{equation}
Therefore, by triangle inequality, since $\mnorm{S^t}=n-r$, we have
\begin{equation}
    \label{eqn:update-bnd-dim}
    \norm{\g^t}\leq\sum_{j\in S^t}\norm{\nabla Q_j(x^t)}\leq (n-r)(2n\mu\epsilon+\mu\norm{x^t-x^*}).
\end{equation}
Substituting from above in \eqref{eqn:apprx-one-step}, we obtain that
\begin{align}
    \label{eqn:rate-1-dim}
    \norm{x^{t+1}-x^*}^2\leq&\norm{x^t-x^*}^2-2\eta_t\iprod{x^t-x^*}{\g^t}+\eta_t^2\norm{\g^t}^2 \nonumber \\
        \leq&\norm{x^t-x^*}^2-2\eta_t\iprod{x^t-x^*}{\g^t}+\eta_t^2(n-r)^2(2n\mu\epsilon+\mu\norm{x^t-x^*})^2
\end{align}
We define
\begin{equation}
    \phi_t=\iprod{x^t-x^*}{\g^t}.
\end{equation}
Following the same argument in the \textbf{second} part of the proof of Theorem~\ref{thm:approx}, by Assumption~\ref{assum:strongly-convex}, 
\begin{align}
    \phi_t&\geq n\gamma\norm{x^t-x^*}^2-r\norm{x^t-x^*}(2n\mu\epsilon+\mu\norm{x^t-x^*}) \tag{\ref{eqn:apprx_phi_t_3}}
\end{align}
Let $e_t$ denote $\norm{x^t-x^*}$, and note that $\eta_t=c/(t+1)$ for all $t$ with some $c>0$. Combining above and \eqref{eqn:rate-1-dim}, we obtain that
\begin{align}
    \label{eqn:et-ineq-1-dim}
    e_{t+1}^2\leq&\left[1+2\eta_t\left(\mu r-\gamma n\right)+\eta_t^2\mu^2(n-r)^2\right]e_t^2 + 4\left[\eta_t\mu\epsilon nr+\eta_t^2\mu^2\epsilon(n-r)^2n\right]e_t + 4\eta_t^2\mu^2\epsilon^2(n-r)^2n^2 \nonumber \\
    \leq&\left[1+2\eta_t\left(\mu r-\gamma n\right)+\eta_t^2\mu^2n^2\right]e_t^2 + 4\left(\eta_t\mu\epsilon n^2+\eta_t^2\mu^2\epsilon n^3\right)e_t + 4\eta_t^2\mu^2\epsilon^2n^4 \nonumber \\
    =&\left[1+\frac{2\left(\mu r-\gamma n\right)c}{t+1}+\frac{\mu^2n^2c^2}{(t+1)^2}\right]e_t^2 + \left[\frac{4\mu\epsilon n^2c}{t+1}+\frac{4\mu^2\epsilon n^3c^2}{(t+1)^2}\right]e_t + \frac{4\mu^2\epsilon^2n^4c^2}{(t+1)^2}.
\end{align}

Let us first define $\alpha=1-\cfrac{r}{n}\cdot\cfrac{\mu}{\gamma}$. Consider an arbitrary set of more than $n-r$ agents $S$, and two points $x,y\in\R^d$, $x\neq y$. By Assumption~\ref{assum:lipschitz}, for every $j\in S$, 
\begin{equation}
    \norm{\nabla Q_j(x)-\nabla Q_j(y)}\leq\mu\norm{x-y}.
\end{equation}
By Cauchy-Schwartz inequality,
\begin{equation}
    \iprod{x-y}{\nabla Q_j(x)-\nabla Q_j(y)}\leq\norm{x-y}\cdot\norm{\nabla Q_j(x)-\nabla Q_j(y)}\leq\mu\norm{x-y}^2.
\end{equation}
Therefore, 
\begin{equation}
    \label{eqn:cmp-mu-dim}
    \sum_{j\in S}\iprod{x-y}{\nabla Q_j(x)-\nabla Q_j(y)}\leq\mu\mnorm{S}\norm{x-y}^2.
\end{equation}
On the other hand, by Assumption~\ref{assum:strongly-convex}, 
\begin{equation}
    \iprod{x-y}{\nabla Q_S(x)-\nabla Q_S(y)}\geq\gamma\norm{x-y}^2.
\end{equation}
Recall that $Q_S(x)=\cfrac{1}{\mnorm{S}}\sum_{j\in S}Q_j(x)$. From above we obtain that
\begin{equation}
    \label{eqn:cmp-gamma-dim}
    \sum_{j\in S}\iprod{x-y}{\nabla Q_j(x)-\nabla Q_j(y)}\geq\gamma\mnorm{S}\norm{x-y}^2.
\end{equation}
Compare \eqref{eqn:cmp-mu-dim} and \eqref{eqn:cmp-gamma-dim}, as $x$ and $y$ are arbitrary and non-identical points, suppose Assumptions~\ref{assum:lipschitz} and \ref{assum:strongly-convex} hold at the same time, we have $\gamma<\mu$. Therefore, 
\begin{equation}
    \frac{\gamma}{\mu}\cdot\alpha=\frac{\gamma}{\mu}-\frac{r}{n}\leq\frac{\gamma}{\mu}\leq1.
    \label{eqn:gamma-mu-alpha-2}
\end{equation}

Let us define 
\begin{equation}
    A(t)=1+\frac{2\left(\mu r-\gamma n\right)c}{t+1}+\frac{\mu^2n^2c^2}{(t+1)^2},
\end{equation} 
the coefficient of the quadratic term in \eqref{eqn:et-ineq-1-dim}. Note that for all $t$, 
\begin{equation}
    A(t)=1+\frac{2\left(\mu r-\gamma n\right)c}{t+1}+\frac{\mu^2n^2c^2}{(t+1)^2}\leq1+{2\left(\mu r-\gamma n\right)c}+{\mu^2n^2c^2}\triangleq A.
\end{equation}
Now, consider the value of $A$. By the definition of $\alpha$, $\mu r-\gamma n=-\gamma n\alpha$. So $A$ can be written as
\begin{equation}
    A=1-2\gamma n \alpha c + \mu^2n^2c^2=1-(\mu n)^2\cdot c\left(\frac{2\gamma\alpha}{\mu^2 n}-c\right).
\end{equation}
Let $\overline{\eta}=\cfrac{2\gamma\alpha}{\mu^2 n}$, from above we obtain that
\begin{equation}
    \label{eqn:def-A-dim}
    A=1-(\mu n)^2c(\overline{\eta}-c).
\end{equation}
Note that
\begin{equation}
    c(\overline{\eta}-c)=\left(\frac{\overline{\eta}}{2}\right)^2-\left(c-\frac{\overline{\eta}}{2}\right)^2.
\end{equation}
Therefore, 
\begin{align}
    A=&1-(\mu n)^2\left[\left(\frac{\overline{\eta}}{2}\right)^2-\left(c-\frac{\overline{\eta}}{2}\right)^2\right] \nonumber \\
    =&(\mu n)^2\left(c-\frac{\overline{\eta}}{2}\right)^2+1-(\mu n)^2\left(\frac{\overline{\eta}}{2}\right)^2.
\end{align}
Since $c\in(0,\overline{\eta})$, the minimum value of $A$ can be obtained when $c=\overline{\eta}/2$, 
\begin{equation}
    \min_{\eta}A=1-(\mu n)^2\left(\frac{\overline{\eta}}{2}\right)^2.
\end{equation}
On the other hand, since $c\in(0,\overline{\eta})$, from \eqref{eqn:def-A}, $A<1$. Thus,
\begin{equation}
    1-(\mu n)^2\left(\frac{\overline{\eta}}{2}\right)^2\leq A<1.
\end{equation}
Substituting $\overline{\eta}=\cfrac{2\gamma\alpha}{\mu^2 n}$ in above implies that $A\in\left[1-\left(\cfrac{\gamma\alpha}{\mu}\right)^2,1\right)$. Since $\cfrac{\gamma\alpha}{\mu}\leq1$ (cf. \eqref{eqn:gamma-mu-alpha-2}), $A\in[0,1)$.

Let $B=4\mu\epsilon n^2c$, $C=4\mu^2\epsilon n^3c^2$, $D=4\mu^2\epsilon^2 n^4c^2$. The inequality of \ref{eqn:et-ineq-1-dim} can be written as
\begin{equation}
    \label{eqn:et-ineq-2-dim}
    e_{t+1}^2\leq A(t)e_t^2+\left[B/(t+1)+C/(t+1)^2\right]e_t+D/(t+1)^2.
\end{equation}
Recall that $x^t\in\W$ for all $t$, where $\W$ is a compact set. There exists a $\Gamma=\max_{x\in\W}\norm{x-x^*}<\infty$, such that $\norm{x^t-x^*}\leq\Gamma$ for all $t$, or $e_t\leq\Gamma$. Thus, from \eqref{eqn:et-ineq-2-dim} we obtain
\begin{equation}
    \label{eqn:et-ineq-3-dim}
    e_{t+1}^2\leq A(t)e_t^2+\frac{B\Gamma}{t+1}+\frac{C\Gamma+D}{(t+1)^2}.
\end{equation}

Knowing that $A<1$, from \eqref{eqn:et-ineq-3-dim} we obtain by induction that
\begin{align}
    e_{t+1}^2\leq& Ae_t^2+\frac{B\Gamma}{t+1}+\frac{C\Gamma+D}{(t+1)^2} \nonumber \\
        \leq& A\left(Ae_{t-1}^2+\frac{B\Gamma}{t}+\frac{C\Gamma+D}{t^2}\right)+\frac{B\Gamma}{t+1}+\frac{C\Gamma+D}{(t+1)^2} \nonumber \\
        \leq& A^2e_{t-1}^2+B\Gamma\left(\frac{A}{t}+\frac{1}{t+1}\right)+(C\Gamma+D)\left(\frac{A}{t^2}+\frac{1}{(t+1)^2}\right) \nonumber \\
        \leq&... \nonumber \\ 
        \leq& A^{t+1}e_{0}^2+B\Gamma\left(\frac{A^t}{1}+...+\frac{A}{t}+\frac{1}{t+1}\right)+(C\Gamma+D)\left(\frac{A^t}{1^2}+...+\frac{A}{t^2}+\frac{1}{(t+1)^2}\right) \nonumber \\
        \leq& A^{t+1}e_{0}^2+B\Gamma\left(A^t+...+A+1\right)+(C\Gamma+D)\left(A^t+...+A+1\right) \nonumber \\
        \leq& A^{t+1}e_{0}^2+\frac{(B+C)\Gamma+D}{1-A}.
\end{align}
Since $A<1$, $A^t<1$ for all $t$. Thus,
\begin{equation}
    e_{t+1}^2\leq A^{t+1}e_0^2+\frac{(B+C)\Gamma+D}{1-A}.
\end{equation}
Let $R=\cfrac{(B+C)\Gamma+D}{1-A}$, we have
\begin{align}
    \norm{x^t-x^*}^2\leq A^{t}\norm{x^0-x^*}^2+R
\end{align}

Note that since $B=\Theta(\epsilon)$, $C=\Theta(\epsilon)$, and $D=\Theta(\epsilon^2)$, we have $R=\Theta(\epsilon)$.

%% file: appendix-gen.tex
\section{Appendix: Proof for Theorem~\ref{thm:approx-generalized}}
\label{appdx:proof-thm-generalized}

\noindent \fbox{\begin{minipage}{\linewidth}
\textbf{Theorem~\ref{thm:approx-generalized}.}
    \textit{Suppose Assumption~\ref{assum:lipschitz} and \ref{assum:strongly-convex} hold true, and the cost functions of all agents satisfies $(r,\epsilon)$-redundancy. 
     Assume that $\eta_t$ in (\ref{eqn:update-straggler}) satisfies $\sum_{t=0}^\infty\eta_t=\infty$, $\sum_{t=0}^\infty\eta_t^2<\infty$, and $\eta_t\geq\eta_{t+1}$ for all $t$.
     Let $\alpha$ and $D$ be defined as follows: 
     \[\alpha\triangleq1-\dfrac{r}{n}\cdot\dfrac{\mu}{\gamma}>0 ~\textrm{ and }~ D\triangleq\dfrac{2r\mu}{\alpha\gamma}\epsilon.\] 
     Then, suppose there exists a $\tau\geq0$ such that $\mnorm{T^t}\geq n-r$ for all $t$, for the proposed algorithm with update rule \eqref{eqn:update-straggler}, 
    \begin{equation*}
        \lim_{t\rightarrow\infty}\norm{x^t-x^*}\leq\D, 
    \end{equation*}
    where $x^*=\arg\min_x\sum_{j\in[n]}Q_j(x)$.}
\end{minipage}}

The proof consists of three parts. In the first part, we show the norm of the update is bounded for all $t$. In the second part, we consider the inner product term $\phi_{t;1}$ (defined later) is lower-bounded. And in the third part, we show that with a lower-bounded $\phi_{t;1}$ term, the iterative estimate converges. 

\textbf{First}, we show that $\norm{\sum_{i=0}^\tau\sum_{j\in T^{t;t-i}}\nabla Q_j(x^{t-i})}$ is bounded for all $t$. By Assumption~\ref{assum:lipschitz} and $(f,r)$-redundancy, following the same argument in the proof of Theorem~\ref{thm:approx}, we obtain that for all $j\in[n]$,
\begin{equation}
    \norm{\nabla Q_j(x^*)}\leq(2n-2r+1)\mu\epsilon. \tag{\ref{eqn:apprx_gradient_bnd}}
\end{equation}
Furthermore, by Assumption~\ref{assum:lipschitz}, for all $x\in\R^d$,
\begin{equation}
    \norm{\nabla Q_j(x)}\leq 2n\mu\epsilon+\mu\norm{x-x^*}. \tag{\ref{eqn:apprx_gradient_bnd_2}}
\end{equation}
Recall that $x^t\in\W$ for all $t$, where $\W$ is a compact set. There exists a $\Gamma=\max_{x\in\W}\norm{x-x^*}<\infty$, such that $\norm{x^t-x^*}\leq\Gamma$ for all $t$. Therefore, for all $t$,
\begin{align}
    &\norm{\sum_{i=0}^\tau\sum_{j\in T^{t;t-i}}\nabla Q_j(x^{t-i})}\leq \sum_{i=0}^\tau\sum_{j\in T^{t;t-i}}\norm{\nabla Q_j(x^{t-i})} \nonumber \\
    \leq& \sum_{i=0}^\tau\mnorm{T^{t;t-i}}\cdot(2n\mu\epsilon+\mu\norm{x^{t-i}-x^*})\leq \sum_{i=0}^\tau\mnorm{T^{t;t-i}}\cdot(2n\mu\epsilon+\mu\Gamma)
\end{align}
Recall the definition of $T^{t;t-i}$, that $T^{t;t-i_1}\cap T^{t;t-i_2}=\varnothing$ for all $0\leq i_1, i_2\leq\tau$, $i_1\neq i_2$. Therefore, $\sum_{i=0}^\tau\mnorm{T^{t;t-i}}\leq n$. Thus,
\begin{align}
    \label{eqn:update-bnd-generalized}
    &\norm{\sum_{i=0}^\tau\sum_{j\in T^{t;t-i}}\nabla Q_j(x^{t-i})}\leq n(2n\mu\epsilon+\mu\Gamma)<\infty.
\end{align}

\textbf{Second}, consider the term 
\begin{equation}
    \label{eqn:phi_t-gen}
    \phi_t=\iprod{x^t-x^*}{\sum_{i=0}^\tau\sum_{j\in T^{t;t-i}}\nabla Q_j(x^{t-i})}.
\end{equation}
Note that 
\begin{equation}
    \label{eqn:origin-of-phi-t}
    \norm{x^t-x^*+\eta_t\sum_{i=0}^\tau\sum_{j\in T^{t;t-i}}\nabla Q_j(x^{t-i})}\leq\norm{x^t-x^*}-2\eta_t\phi_t+\norm{\sum_{i=0}^\tau\sum_{j\in T^{t;t-i}}\nabla Q_j(x^{t-i})}.
\end{equation}

Let $T^t=\bigcup_{i=0}^\tau T^{t;t-i}$ be the set of agents whose gradients are used for the update at iteration $t$.
\begin{align}
    \sum_{i=0}^\tau\sum_{j\in T^{t;t-i}}\nabla Q_j(x^{t-i})=&\sum_{i=0}^\tau\sum_{j\in T^{t;t-i}}\left(\nabla Q_j(x^t) - \nabla Q_j(x^t) + \nabla Q_j(x^{t-i}) \right) \nonumber \\
    =&\sum_{i=0}^\tau\sum_{j\in T^{t;t-i}}\nabla Q_j(x^t) - \sum_{i=0}^\tau\sum_{j\in T^{t;t-i}}\left(\nabla Q_j(x^{t}) - \nabla Q_j(x^{t-i})\right) \nonumber \\
    =&\sum_{j\in T^t}\nabla Q_j(x^t) - \sum_{i=0}^\tau\sum_{j\in T^{t;t-i}}\left(\nabla Q_j(x^{t}) - \nabla Q_j(x^{t-i})\right).
\end{align}
Therefore, we have
\begin{align}
    \label{eqn:phi_t-generalized}
    \phi_t=&\iprod{x^t-x^*}{\sum_{i=0}^\tau\sum_{j\in T^{t;t-i}}\nabla Q_j(x^{t-i})} \nonumber \\
    =&\iprod{x^t-x^*}{\sum_{j\in T^{t}}\nabla Q_j(x^{t})} - \iprod{x^t-x^*}{\sum_{i=0}^\tau\sum_{j\in T^{t;t-i}}\left(\nabla Q_j(x^t)-\nabla Q_j(x^{t-i})\right)}.
\end{align}

For the first term in \eqref{eqn:phi_t-generalized}, denote it $\phi_{t;1}$, we have
\begin{align}
    \label{eqn:phi_t1}
    \phi_{t;1}=&\iprod{x^t-x^*}{\sum_{j\in T^{t}}\nabla Q_j(x^{t})} \nonumber \\
    =&\iprod{x^t-x^*}{\sum_{j\in T^{t}}\nabla Q_j(x^{t})+\sum_{k\in[n]\backslash T^t}\nabla Q_k(x^t)-\sum_{k\in[n]\backslash T^t}\nabla Q_k(x^t)} \nonumber \\
    =&\iprod{x^t-x^*}{\sum_{j\in[n]}\nabla Q_j(x^{t})}-\iprod{x^t-x^*}{\sum_{k\in[n]\backslash T^t}\nabla Q_k(x^t)}.
\end{align}
Recall \eqref{eqn:apprx_phi_t_1_1} and \eqref{eqn:apprx_phi_t_1_2}, we have the same result as we have in \eqref{eqn:apprx_phi_t_2}:
\begin{equation}
    \phi_{t;1}\geq n\gamma\norm{x^t-x^*}^2-\sum_{k\in[n]\backslash T^t}\norm{x^t-x^*}\norm{\nabla Q_k(x^t)}.
\end{equation}
Note that $\mnorm{T^t}\geq n-r$, or $\mnorm{[n]\backslash T^t}\leq r$. Combining above and \eqref{eqn:apprx_gradient_bnd_2},
\begin{align}
    \label{eqn:phi_t1_bnd}
    \phi_{t;1}\geq& n\gamma\norm{x^t-x^*}^2-r\norm{x^t-x^*}\left(2n\mu\epsilon+\mu\norm{x-x^*}\right) \nonumber \\
        =&n\gamma\left(1-\dfrac{r}{n}\cdot\dfrac{\mu}{\gamma}\right)\norm{x^t-x^*}\left(\norm{x^t-x^*}-\dfrac{2r\mu\epsilon}{\gamma\left(1-\dfrac{r}{n}\cdot\dfrac{\mu}{\gamma}\right)}\right).
\end{align}
Recall that we assume $\displaystyle\alpha=1-\dfrac{r}{n}\cdot\dfrac{\mu}{\gamma}>0$.
We have
\begin{equation}
    \phi_{t;1}\geq\alpha n\gamma\norm{x^t-x^*}\left(\norm{x^t-x^*}-\dfrac{2r\mu}{\alpha\gamma}\epsilon\right).
    \label{eqn:phi_t1_final}
\end{equation}
Let $\D\triangleq\dfrac{2r\mu}{\alpha\gamma}\epsilon$. \eqref{eqn:phi_t1_final} implies for an arbitrary $\delta>0$, 
\begin{equation}
    \label{eqn:phi_t1_final_final}
    \phi_{t;1}\geq\alpha n\gamma\delta\left(D+\delta\right)>0~\textrm{ when }~\norm{x^t-x^*}\geq\D+\delta. 
\end{equation}

For the second term in \eqref{eqn:phi_t-generalized}, denote it $\phi_{t;2}$, by Cauchy-Schwartz inequality, we have
\begin{align}
    \label{eqn:phi_t2}
    \phi_{t;2}=&\iprod{x^t-x^*}{\sum_{i=0}^\tau\sum_{j\in T^{t;t-i}}\left(\nabla Q_j(x^{t}) - \nabla Q_j(x^{t-i})\right)} \nonumber \\
    \leq& \norm{x^t-x^*}\cdot\norm{\sum_{i=0}^\tau\sum_{j\in T^{t;t-i}}\left(\nabla Q_j(x^{t}) - \nabla Q_j(x^{t-i})\right)}.
\end{align}
Consider the factor $\norm{\sum_{i=0}^t\sum_{j\in T^{t;t-i}}\left(\nabla Q_j(x^{t}) - \nabla Q_j(x^{t-i})\right)}$. By triangle inequality,
\begin{align}
    \label{eqn:phi_t2_step_a}
    \norm{\sum_{i=0}^\tau\sum_{j\in T^{t;t-i}}\left(\nabla Q_j(x^{t}) - \nabla Q_j(x^{t-i})\right)}\leq&\sum_{i=0}^\tau\sum_{j\in T^{t;t-i}}\norm{\nabla Q_j(x^{t}) - \nabla Q_j(x^{t-i})}.
\end{align}
By Assumption~\ref{assum:lipschitz} and $x^t\in\W$ for all $t$, 
\begin{equation}
    \label{eqn:phi_t2_step_b}
    \norm{\nabla Q_j(x^{t}) - \nabla Q_j(x^{t-i})}\leq\mu\norm{x^t-x^{t-i}}.
\end{equation}
According to the update rule \eqref{eqn:update-straggler}, 
\begin{align}
    x^{t}&=x^{t-1}-\eta_{t-1}\sum_{h=0}^\tau\sum_{j\in T^{t-1;t-1-h}}\nabla Q_j(x^{t-1-h}) \nonumber \\
        &=x^{t-2}-\eta_{t-1}\sum_{h=0}^\tau\sum_{j\in T^{t-1;t-1-h}}\nabla Q_j(x^{t-1-h})-\eta_{t-2}\sum_{h=0}^\tau\sum_{j\in T^{t-2;t-2-h}}\nabla Q_j(x^{t-2-h}) \nonumber \\
        &=x^{t-2}-\sum_{k=1}^2\eta_{t-k}\sum_{h=0}^\tau\sum_{j\in T^{t-k;t-k-h}}\nabla Q_j(x^{t-k-h}) \nonumber \\
        &=\cdots \nonumber \\
        &=x^{t-i}-\sum_{k=1}^i\eta_{t-k}\sum_{h=0}^\tau\sum_{j\in T^{t-k;t-k-h}}\nabla Q_j(x^{t-k-h}).
\end{align}
Therefore,
\begin{align}
    \norm{x^t-x^{t-i}}=&\norm{\sum_{k=1}^i\eta_{t-k}\sum_{h=0}^\tau\sum_{j\in T^{t-k;t-k-h}}\nabla Q_j(x^{t-k-h})} \nonumber \\
        \leq&\sum_{k=1}^i\eta_{t-k}\norm{\sum_{h=0}^\tau\sum_{j\in T^{t-k;t-k-h}}\nabla Q_j(x^{t-k-h})}.
\end{align}
By \eqref{eqn:update-bnd-generalized}, let $\overline{G}=n(2n\mu\epsilon+\mu\Gamma)$. Also note that for all $t$, $\eta_{t}\geq\eta_{t+1}$. From above we have
\begin{equation}
    \norm{x^t-x^{t-i}}\leq\sum_{k=1}^i\eta_{t-k}\overline{G}\leq i\cdot \eta_{t-i}\overline{G}. 
\end{equation}
Combining \eqref{eqn:phi_t2_step_a}, \eqref{eqn:phi_t2_step_b} and above, we have
\begin{align}
    &\norm{\sum_{i=0}^\tau\sum_{j\in T^{t;t-i}}\left(\nabla Q_j(x^{t}) - \nabla Q_j(x^{t-i})\right)}\leq\sum_{i=0}^\tau\sum_{j\in T^{t;t-i}}\norm{\nabla Q_j(x^{t}) - \nabla Q_j(x^{t-i})} \nonumber \\
        \leq&\sum_{i=0}^\tau\sum_{j\in T^{t;t-i}}\mu i\cdot \eta_{t-i}\overline{G} \leq\overline{G}\cdot\sum_{i=0}^\tau\sum_{j\in T^{t;t-i}}\mu\tau\eta_{t-\tau}=\overline{G}\cdot\sum_{j\in T^{t}}\mu\tau\eta_{t-\tau} \nonumber \\
        \leq&\mu\tau\eta_{t-\tau}n\overline{G}.
\end{align}
Combining with \eqref{eqn:phi_t2}, from above we have
\begin{equation}
    \label{eqn:phi_t2_bnd}
    \phi_{t;2}\leq\mu\tau\eta_{t-\tau}n\overline{G}\norm{x^t-x^*}.
\end{equation}
In summary, $\phi_t=\phi_{t;1}-\phi_{t;2}$. 

\textbf{Third}\footnote{Note that this part of the proof is similar to what we have in the proof of Lemma~\ref{lemma:bound} in Appendix~\ref{appdx:proof-lemma-bound}. Still, the full argument are presented with repeated contents to avoid confusion.}, we show that assuming that there exists some $\D^*\in[0,\max_{x\in\W}\norm{x-x^*})$ and $\xi>0$, such that $\phi_{t;1}\geq\xi$ when $\norm{x^t-x^*}\geq\D^*$, we have $\lim_{t\rightarrow\infty}\norm{x^t-x^*}\leq\D^*$.

Let $e^t$ denote $\norm{x^t-x^*}$. Define a scalar function $\psi$,
\begin{equation}
    \psi(y)=\left\{\begin{array}{ll}
        0, & y<\left(\D^*\right)^2, \\
        \left(y-\left(\D^*\right)^2\right)^2, & \textrm{otherwise}.
    \end{array}\right.
    \label{eqn:psi_def-gen}
\end{equation}
Let $\psi'(y)$ denote the derivative of $\psi$ at $y$. Then (cf. \cite{bottou1998online})
\begin{equation}
    \label{eqn:psi_bnd-gen}
    \psi(z)-\psi(y)\leq(z-y)\psi'(y)+(z-y)^2,~\forall y,z\in\mathbb{R}_{\geq0}.
\end{equation}
Note,
\begin{equation}
    \label{eqn:psi_prime-gen}
    \psi'(y)=\max\left\{0,2\left(y-\left(\D^*\right)^2\right)\right\}.
\end{equation}

Now, define
\begin{equation}
    \label{eqn:ht_def-gen}
    h_t\triangleq\psi(e_t^2).
\end{equation}
From \eqref{eqn:psi_bnd-gen} and \eqref{eqn:ht_def-gen},
\begin{equation}
    h_{t+1}-h_t=\psi\left(e_{t+1}^2\right)-\psi\left(e_t^2\right)\leq\left(e_{t+1}^2-e_t^2\right)\psi'\left(e_t^2\right)+\left(e_{t+1}^2-e_t^2\right)^2,~\forall t\in\mathbb{Z}_{\geq0}.
\end{equation}
From now on, we use $\psi_t'$ as the shorthand for $\psi'\left(e_t^2\right)$, i.e.,
\begin{equation}
    \label{eqn:def-psi-t-gen}
    \psi_t'\triangleq\psi'\left(e_t^2\right).
\end{equation}
From above, for all $t\geq0$,
\begin{equation}
    \label{eqn:ht_1-gen}
    h_{t+1}-h_t\leq\left(e_{t+1}^2-e_t^2\right)\psi'_t+\left(e_{t+1}^2-e_t^2\right)^2.
\end{equation}

Recall the iterative process \eqref{eqn:update-straggler}. From now on, we use $\mathsf{StragAgg}[t]$ as a short hand for the output of the gradient aggregation rule at iteration $t$, i.e. 
\begin{equation}
    \label{eqn:def-stragg-t}
    \mathsf{StragAgg}[t]\triangleq\sum_{i=0}^\tau\sum_{j\in T^{t;t-i}}\nabla Q_j(x^{t-i}).
\end{equation}
Using the non-expansion property of Euclidean projection onto a closed convex set\footnote{$\norm{x-x^*}\geq\norm{[x]_\W-x^*},~\forall w\in\mathbb{R}^d$.}, 
\begin{equation}
    \norm{x^{t+1}-x^*}\leq\norm{x^t-\eta_t\mathsf{StragAgg}[t]-x^*}.
\end{equation}
Taking square on both sides,
\begin{equation*}
    e_{t+1}^2\leq e_t^2-2\eta_t\iprod{x_t-x^*}{\mathsf{StragAgg}[t]}+\eta_t^2\norm{\mathsf{StragAgg}[t]}^2.
\end{equation*}
Recall from \eqref{eqn:phi_t-gen} that $\iprod{x_t-x^*}{\mathsf{StragAgg}[t]}=\phi_t$, therefore (cf. \eqref{eqn:origin-of-phi-t}),
\begin{equation}
    \label{eqn:proj_bound-generalized}
    e_{t+1}^2\leq e_t^2-2\eta_t\phi_t+\eta_t^2\norm{\mathsf{StragAgg}[t]}^2,~\forall t\geq0.
\end{equation}

As $\psi'_t\geq0,~\forall t\in\mathbb{Z}_{\geq0}$, combining \eqref{eqn:ht_1-gen} and \eqref{eqn:proj_bound-generalized},
\begin{equation}
    \label{eqn:ht_2-gen}
    h_{t+1}-h_t\leq\left(-2\eta_t\phi_t+\eta_t^2\norm{\mathsf{StragAgg}[t]}^2\right)\psi'_t+\left(e_{t+1}^2-e_t^2\right)^2,~\forall t\geq0.
\end{equation}
Note that 
\begin{equation}
    \mnorm{e_{t+1}^2-e_t^2}=(e_{t+1}+e_t)\mnorm{e_{t+1}-e_t}.
\end{equation}
As $\W$ is assumed compact, there exists
\begin{equation}
    \Gamma=\max_{x\in\W}\norm{x-x^*}\leq\infty.
\end{equation}
Let $\Gamma>0$, since otherwise $\W=\{x^*\}$ only contains one point, and the problem becomes trivial. As $x^t\in\W$, $\forall t\geq0$,
\begin{equation}
    \label{eqn:e_t_bound-gen}
    e_t\leq\Gamma,
\end{equation}
which implies
\begin{equation}
    e_{t+1}+e_t\leq2\Gamma.
\end{equation}
Therefore,
\begin{equation}
    \mnorm{e_{t+1}^2-e_t^2}\leq2\Gamma\mnorm{e_{t+1}-e_t},~\forall t\geq0.
    \label{eqn:e2t-bound-1-gen}
\end{equation}
By triangle inequality,
\begin{equation}
    \mnorm{e_{t+1}-e_t}=\mnorm{\norm{x^{t+1}-x^*}-\norm{x^t-x^*}}\leq\norm{x^{t+1}-x^t}.
    \label{eqn:e2t-bound-2-gen}
\end{equation}
From \eqref{eqn:update-straggler} and the non-expansion property of Euclidean projection onto a closed convex set,
\begin{equation}
    \norm{x^{t+1}-x^t}=\norm{\left[x^t-\eta_t\mathsf{StragAgg}[t]\right]_\W-x^t}\leq\eta_t\norm{\mathsf{StragAgg}[t]}.
    \label{eqn:e2t-bound-3-gen}
\end{equation}
So from \eqref{eqn:e2t-bound-1-gen}, \eqref{eqn:e2t-bound-2-gen}, and \eqref{eqn:e2t-bound-3-gen},
\begin{align}
    &\mnorm{e_{t+1}^2-e_t^2}\leq2\eta_t\Gamma\norm{\mathsf{StragAgg}[t]}, \nonumber \\
    \Longrightarrow&\left(e_{t+1}^2-e_t^2\right)^2\leq4\eta_t^2\Gamma^2\norm{\mathsf{StragAgg}[t]}^2.
\end{align}
Substituting above in \eqref{eqn:ht_2-gen},
\begin{align}
    h_{t+1}-h_t\leq&\left(-2\eta_t\phi_t+\eta_t^2\norm{\mathsf{StragAgg}[t]}^2\right)\psi'_t+4\eta_t^2\Gamma^2\norm{\mathsf{StragAgg}[t]}^2, \nonumber \\
    =&-2\eta_t\phi_t\psi'_t+(\psi'_t+4\Gamma^2)\eta_t^2\norm{\mathsf{StragAgg}[t]}^2,~\forall t\geq0.
\end{align}
Recall that $\phi_t=\phi_{t;1}-\phi_{t;2}$. Also, $\eta_t\geq\eta_{t+1}$ for all $t$. Substitute for \eqref{eqn:e_t_bound-gen} and \eqref{eqn:phi_t2_bnd} in above, we obtain that
\begin{align}
    \label{eqn:ht_3-gen}
    h_{t+1}-h_t\leq&-2\eta_t(\phi_{t;1}-\phi_{t;2})\psi'_t+(\psi'_t+4\Gamma^2)\eta_t^2\norm{\mathsf{StragAgg}[t]}^2,~\forall t\geq0 \nonumber \\
    \leq&-2\eta_t\phi_{t;1}\psi'_t+2\eta_t(\mu\tau\eta_{t-\tau}n\overline{G}e_t)\psi'_t+(\psi'_t+4\Gamma^2)\eta_t^2\norm{\mathsf{StragAgg}[t]}^2,~\forall t\geq0 \nonumber \\
    \leq&-2\eta_t\phi_{t;1}\psi'_t+2\eta_t^2\mu\tau n\overline{G}\Gamma\psi'_t+(\psi'_t+4\Gamma^2)\eta_t^2\norm{\mathsf{StragAgg}[t]}^2,~\forall t\geq0. 
\end{align}

Let us assume for now that $\D^*\in[0,\max_{x\in\W}\norm{x-x^*})$, which indicates $\D^*<\Gamma$. Using \eqref{eqn:psi_prime-gen} and \eqref{eqn:e_t_bound-gen}, we have
\begin{equation}
    0\leq\psi'_t\leq2\left(e_t^2-\left(\D^*\right)^2\right)\leq2\left(\Gamma^2-\left(\D^*\right)^2\right)\leq2\Gamma^2.
    \label{eqn:psi_prime_t_bnd-gen}
\end{equation}
From \eqref{eqn:update-bnd-generalized} (recall that $\overline{G}=n(2n\mu\epsilon+\mu\Gamma)$), $\norm{\mathsf{StragAgg}[t]}\leq\overline{G}<\infty$ for all $t$. Substituting \eqref{eqn:psi_prime_t_bnd-gen} in \eqref{eqn:ht_3-gen},
\begin{align}
    \label{eqn:ht_4-gen}
    h_{t+1}-h_t\leq&-2\eta_t\phi_{t;1}\psi'_t+2\eta_t^2\mu\tau n\overline{G}\Gamma\left(2\Gamma^2\right)+\left(2\Gamma^2+4\Gamma^2\right)\eta_t^2\overline{G}^2 \nonumber \\
    =&-2\eta_t\phi_{t;1}\psi'_t+\eta_t^2\left(4\mu\tau n\Gamma^3\overline{G}+6\Gamma^2\overline{G}^2\right),~\forall t\geq0.
\end{align}

Now we use Lemma~\ref{lemma:converge} to show that $h_\infty=0$ as follows. For each iteration $t$, consider the following two cases:
\begin{description}
    \item[Case 1)] Suppose $e_t<\D^*$. In this case, $\psi'_t=0$. By Cauchy-Schwartz inequality,
        \begin{equation}
            \mnorm{\phi_{t;1}}=\mnorm{\iprod{x^t-x^*}{\mathsf{StragAgg}[t]}}\leq e_t\norm{\mathsf{StragAgg}[t]}.
        \end{equation}
        By \eqref{eqn:update-bnd-generalized} and \eqref{eqn:e_t_bound-gen}, this implies that
        \begin{equation}
            \mnorm{\phi_{t;1}}\leq\Gamma\overline{G}<\infty.
        \end{equation}
        Thus,
        \begin{equation}
            \label{eqn:phitpsit_1-gen}
            \phi_{t;1}\psi'_t=0.
        \end{equation}
    \item[Case 2)] Suppose $e_t\geq\D^*$. Therefore, there exists $\delta\geq0$, $e_t=\D^*+\delta$. From \eqref{eqn:psi_prime-gen}, we obtain that
    \begin{equation}
        \psi_t'=2\left(\left(\D^*+\delta\right)^2-\left(\D^*\right)^2\right)=2\delta\left(2\D^*+\delta\right).
    \end{equation}
    The statement of Lemma~\ref{lemma:bound} assumes that $\phi_t\geq\xi>0$ when $e_t\geq\D^*$, thus, 
        \begin{equation}
            \label{eqn:phitpsit_2-gen}
            \phi_{t;1}\psi'_t\geq2\delta\xi\left(2\D^*+\delta\right)>0.
        \end{equation}
\end{description}
From \eqref{eqn:phitpsit_1-gen} and \eqref{eqn:phitpsit_2-gen}, for both cases,
\begin{equation}
    \phi_{t;1}\psi'_t\geq0, ~\forall t\geq0.
    \label{eqn:phi_psi_bound-gen}
\end{equation}
Combining this with \eqref{eqn:ht_4-gen},
\begin{equation}
    h_{t+1}-h_t\leq\eta_t^2\left(4\mu\tau n\Gamma^3\overline{G}+6\Gamma^2\overline{G}^2\right).
\end{equation}
From above we have
\begin{equation}
    \left(h_{t+1}-h_t\right)_+\leq\eta_t^2\left(4\mu\tau n\Gamma^3\overline{G}+6\Gamma^2\overline{G}^2\right).
\end{equation}
Since $\sum_{t=0}^\infty\eta_t^2<\infty$, $\Gamma,\overline{G}<\infty$,
\begin{equation}
    \sum_{t=0}^\infty\left(h_{t+1}-h_t\right)_+\leq\left(4\mu\tau n\Gamma^3\overline{G}+6\Gamma^2\overline{G}^2\right)\sum_{t=0}^\infty\eta_t^2<\infty.
\end{equation}
Then Lemma~\ref{lemma:converge} implies that by the definition of $h_t$, we have $h_t\geq0,~\forall t$, 
\begin{align}
    &h_t\xrightarrow[t\rightarrow\infty]{}h_\infty<\infty,~\textrm{and} \label{eqn:upper_bound_h_infty-gen}\\
    &\sum_{t=0}^\infty\left(h_{t+1}-h_t\right)_->-\infty.
\end{align}
Note that $h_\infty-h_0=\sum_{t=0}^\infty(h_{t+1}-h_t)$. Thus, from \eqref{eqn:ht_4-gen} we have 
\begin{equation}
    h_\infty-h_0\leq-2\sum_{t=0}^\infty\eta_t\phi_{t;1}\psi_t'+\left(4\mu\tau n\Gamma^3\overline{G}+6\Gamma^2\overline{G}^2\right)\sum_{t=0}^\infty\eta_t^2.
\end{equation}
By \eqref{eqn:ht_def} the definition of $h_t$, $h_t\geq0$ for all $t$. Therefore, from above we obtain
\begin{equation}
    2\sum_{t=0}^\infty\eta_t\phi_{t;1}\psi_t'\leq h_0-h_\infty+\left(4\mu\tau n\Gamma^3\overline{G}+6\Gamma^2\overline{G}^2\right)\sum_{t=0}^\infty\eta_t^2.
    \label{eqn:bound_2_sum-gen}
\end{equation}
By assumption, $\sum_{t=0}^\infty\eta_t^2<\infty$. Substituting from \eqref{eqn:e_t_bound-gen} that $e_t<\infty$ in \eqref{eqn:ht_def-gen}, we obtain that 
\begin{equation}
    h_0=\psi\left(e_0^2\right)<\infty.
\end{equation}
Therefore, \eqref{eqn:bound_2_sum-gen} implies that
\begin{equation}
    2\sum_{t=0}^\infty\eta_t\phi_{t;1}\psi_t'\leq h_0+\left(4\mu\tau n\Gamma^3\overline{G}+6\Gamma^2\overline{G}^2\right)\sum_{t=0}^\infty\eta_t^2<\infty.
\end{equation}
Or simply,
\begin{equation}
    \sum_{t=0}^\infty\eta_t\phi_{t;1}\psi_t'<\infty.
    \label{eqn:upper_bound_etatphitpsit-gen}
\end{equation}

Finally, we reason below by contradiction that $h_\infty=0$. Note that for any $\zeta>0$, there exists a unique positive value $\beta$ such that $\zeta=2\beta\left(2\D^*+\sqrt{\beta}\right)^2$. Suppose that $h_\infty=2\beta\left(2\D^*+\sqrt{\beta}\right)^2$ for some positive value $\beta$. As the sequence $\{h_t\}_{t=0}^\infty$ converges to $h_\infty$ (see \eqref{eqn:upper_bound_h_infty-gen}), there exists some finite $\tau\in\Z_{\geq0}$ such that for all $t\geq\tau$, 
\begin{align}
    &\mnorm{h_t-h_\infty}\leq\beta\left(2\D^*+\sqrt{\beta}\right)^2 \\
    \Longrightarrow & h_t\geq h_\infty-\beta\left(2\D^*+\sqrt{\beta}\right)^2.
\end{align}
As $h_\infty=2\beta\left(2\D^*+\sqrt{\beta}\right)^2$, the above implies that
\begin{equation}
    h_t\geq \beta\left(2\D^*+\sqrt{\beta}\right)^2, \forall t\geq\tau.
    \label{eqn:ht_lower_bound-gen}
\end{equation}
Therefore (cf. \eqref{eqn:psi_def-gen} and \eqref{eqn:ht_def-gen}), for all $t\geq\tau$,
\begin{eqnarray*}
    \left(e_t^2-\left(\D^*\right)^2\right)^2\geq\beta\left(2\D^*+\sqrt{\beta}\right)^2, \textrm{ or} \\
    \mnorm{e_t^2-\left(\D^*\right)^2}\geq\sqrt{\beta}\left(2\D^*+\sqrt{\beta}\right).
\end{eqnarray*}
Thus, for each $t\geq\tau$, either
\begin{equation}
    e^2_t\geq\left(\D^*\right)^2+\sqrt{\beta}\left(2\D^*+\sqrt{\beta}\right)=\left(\D^*+\sqrt{\beta}\right)^2,
    \label{eqn:et_case_1-gen}
\end{equation}
or
\begin{equation}
    e^2_t\leq\left(\D^*\right)^2-\sqrt{\beta}\left(2\D^*+\sqrt{\beta}\right)<\left(\D^*\right)^2.
    \label{eqn:et_case_2-gen}
\end{equation}
If the latter, i.e., \eqref{eqn:et_case_2-gen} holds true for some $t'\geq\tau$, 
\begin{equation}
    h_{t'}=\psi\left(e_{t'}^2\right)=0,
\end{equation}
which contradicts \eqref{eqn:ht_lower_bound-gen}. Therefore, \eqref{eqn:ht_lower_bound-gen} implies \eqref{eqn:et_case_1-gen}.

From above we obtain that if $h_\infty=2\beta\left(2\D^*+\sqrt{\beta}\right)^2$, there exists $\tau<\infty$ such that for all $t\geq\tau$, 
\begin{equation}
    e_t\geq\D^*+\sqrt{\beta}.
\end{equation}
Thus, from \eqref{eqn:phitpsit_2-gen}, with $\delta=\sqrt{\beta}$, we obtain that 
\begin{equation}
    \phi_t\psi_t'\geq2\xi\sqrt{\beta}\left(2\D^*+\sqrt{\beta}\right), \forall t\geq\tau.
\end{equation}
Therefore,
\begin{equation}
    \sum_{t=\tau}^\infty\eta_t\phi_t\psi_t'\geq2\xi\sqrt{\beta}\left(2\D^*+\sqrt{\beta}\right)\sum_{t=\tau}^\infty\eta_t=\infty.
\end{equation}
This is a contradiction to \eqref{eqn:upper_bound_etatphitpsit-gen}. Therefore, $h_\infty=0$, and by \eqref{eqn:ht_def-gen}, the definition of $h_t$, 
\begin{equation}
    h_\infty=\lim_{t\rightarrow\infty}\psi\left(e_t^2\right)=0.
\end{equation}
Hence, by \eqref{eqn:psi_def-gen}, the definition of $\psi(\cdot)$, 
\begin{equation}
    \lim_{t\rightarrow\infty}\norm{x^t-x^*}\leq\D^*.
\end{equation}

\textbf{Finally}, recall the result in \eqref{eqn:phi_t1_final_final}, for arbitrary $\delta>0$,
\begin{equation}
    \tag{\ref{eqn:phi_t1_final_final}}
    \phi_{t;1}\geq\alpha n\gamma\delta\left(D+\delta\right)>0~\textrm{ when }~\norm{x^t-x^*}\geq\D+\delta,
\end{equation}
with $\displaystyle\alpha=1-\dfrac{r}{n}\cdot\dfrac{\mu}{\gamma}>0$ and $\D=\dfrac{2r\mu}{\alpha\gamma}\epsilon$. Combining with the \textbf{third} part of this proof, $\lim_{t\rightarrow\infty}\norm{x^t-x^*}\leq\D$.

%% file: appendix-cge.tex
\section{Appendix: Proof of Theorem~\ref{thm:async-fault-toler-cge}}
\label{appdx:proof-thm-cge}

\noindent \fbox{\begin{minipage}{\linewidth}
    \textbf{Theorem~\ref{thm:async-fault-toler-cge}.}
    \textit{Suppose that Assumptions~\ref{assum:lipschitz} and \ref{assum:strongly-convex} 
    hold true, and the cost functions of all agents satisfies $(f,r;\epsilon)$-redundancy. Assume that $\eta_t$ 
    satisfies $\sum_{t=0}^\infty=\infty$ and $\sum_{t=0}^\infty\eta_t^2<\infty$. If the following conditions holds true: 
    \begin{enumerate}[label=(\roman*)]
        \item $\norm{\mathsf{GradFilter}\left(n-r,f;\,\left\{g_i^t\right\}_{i\in S^t}\right)}<\infty$ for all $t$, and 
        \item if 
        \begin{equation*}
            \displaystyle\alpha = 1 - \frac{f-r}{n-r} + \frac{2\mu}{\gamma}\cdot\frac{f+r}{n-r}>0,
        \end{equation*}
        then for each set of $n-f$ non-faulty agents $\H$, for an arbitrary $\delta > 0$,
        \begin{align*}
            \displaystyle\phi_t \geq \alpha m \gamma \delta \left( \frac{4\mu (f+r)\epsilon}{\alpha\gamma} + \delta \right) > 0~\textrm{ when }~\displaystyle\norm{x^t-x_{\H}}\geq\frac{4\mu (f+r)\epsilon}{\alpha\gamma}+\delta.
        \end{align*}
    \end{enumerate}
    Then for the proposed algorithm with aggregation rule \eqref{eqn:aggregation-rule-ft}, we have \[\lim_{t\rightarrow\infty}\norm{x^t-x_\H}\leq\cfrac{4\mu (f+r)\epsilon}{\alpha\gamma}.\]}
\end{minipage}}

Throughout, we assume $f > 0$ to ignore the trivial case of $f = 0$. 

The gradient filter CGE \cite{liu2021approximate} can be defined as a function on $\R^{d\times m}\rightarrow\R^d$, with two hyperparameters $m$ and $f$. The server receives $m$ gradients $\left\{g_i^t\right\}_{i\in S^t}$ from $m$ agents in the set $S^t$ in iteration t. The server sorts the gradients as per their Euclidean norms (ties broken arbitrarily):
\begin{equation*}
    \norm{g_{i_1}^t}\leq \ldots \leq \norm{g_{i_{m-f}}^t} \leq \norm{g_{i_{m-f+1}}^t}\leq \ldots \leq \norm{g_{i_m}^t}.
\end{equation*}
That is, the gradient with the smallest norm, $g_{i_1}^t$, is received from agent $i_1$, and the gradient with the largest norm, $g_{i_m}^t$, is received from agent $i_m$, with $i_j\in S^t$ for all $j$. Then, the output of the CGE gradient-filter is the vector sum of the $m-f$ gradients with smallest $m-f$ Euclidean norms. Specifically,
\begin{align}
    \gf\left(m,f;\,\left\{g^t_i\right\}_{i\in S^t} \right) = \sum_{j=1}^{m-f}g_{i_j}^t. \label{eqn:cge_gf}
\end{align}
With this definition, we now provide the proof of Theorem~\ref{thm:async-fault-toler-cge}. We set $m=n-r$.

Consider an arbitrary set $\H$ of non-faulty agents with $\mnorm{\H} = n-f$. Recall that under Assumptions~\ref{assum:strongly-convex-ft}, the aggregate cost function $\sum_{i \in \H} Q_i(x)$ has a unique minimum point in set $\W$, which we denote by $x_{\H}$. Specifically, 
\begin{align}
    \left\{x_{\H} \right\} = \W \cap \arg \min_{x \in \R^d} \sum_{i \in \H} Q_i(x). \label{eqn:x_H_W}
\end{align}

\textbf{First}, we show that $\norm{\gf\left(m,f;\,\left\{g^t_i\right\}_{i\in S^t} \right)}=\norm{\sum_{j=1}^{m-f}g_{i_j}^t} < \infty, ~ \forall t$. Consider a subset $S_1 \subset \H$ with $\mnorm{S_1} = m-2f$. From triangle inequality,
\begin{align*}
    \norm{\sum_{j\in S_1}\nabla Q_j(x)-\sum_{j\in S_1}\nabla Q_j(x_{\H})} \leq \sum_{j \in S_1}\norm{\nabla Q_j(x) - \nabla Q_j(x_{\H})}, \quad \forall x \in \R^d.
\end{align*}
Under Assumption~\ref{assum:lipschitz-ft}, i.e., Lipschitz continuity of non-faulty gradients, for each non-faulty agent $j$, $\norm{\nabla Q_j(x) - \nabla Q_j(x_{\H})} \leq \mu \norm{x - x_{\H}}$. Substituting this above implies that 
\begin{equation}
    \norm{\sum_{j\in S_1}\nabla Q_j(x)-\sum_{j\in S_1}\nabla Q_j(x_{\H})} \leq \mnorm{S_1} \mu \, \norm{x-x_{\H}}.
    \label{eqn:lipschitz-distance}
\end{equation}
As $\mnorm{S_1} = m-2f$, the $(f,r;\epsilon)$-redundancy property defined in Definition~\ref{def:approx_redundancy-ft} implies that for all $x_1\in\arg\min_x\sum_{j\in S_1}Q_j(x)$,
\[\norm{x_1-x_{\H}} \leq \epsilon.\]
Substituting from above in~\eqref{eqn:lipschitz-distance} implies that, for all $x_1\in\arg\min_x\sum_{j\in S_1}Q_j(x)$,
\begin{equation}
    \norm{\sum_{j\in S_1}\nabla Q_j(x_1)-\sum_{j\in S_1}\nabla Q_j(x_{\H})} \leq \mnorm{S_1} \mu \, \norm{x_1-x_{\H}} \leq \mnorm{S_1} \mu \epsilon.
    \label{eqn:lipschitz-distance-2}
\end{equation}
For all $x_1\in\arg\min_x\sum_{j\in S_1}Q_j(x)$, $\nabla Q_j(x_1) = 0$. Thus,~\eqref{eqn:lipschitz-distance-2} implies that
\begin{equation}
    \norm{\sum_{j\in S_1}\nabla Q_j(x_{\H})} \leq \mnorm{S_1} \mu \epsilon.
    \label{eqn:lipschitz-distance-S1}
\end{equation}
Now, consider an arbitrary non-faulty agent $i \in \H\setminus S_1$. Let $S_2=S_1\cup\{i\}$. 
Using similar arguments as above we obtain that under the $(f,r;\epsilon)$-redundancy property and Assumption~\ref{assum:lipschitz-ft}, for all $x_2\in\arg\min_x\sum_{j\in S_2}Q_j(x)$,
\begin{align}
    \norm{\sum_{j\in S_2}\nabla Q_j(x_{\H})}=\norm{\sum_{j\in S_2}\nabla Q_j(x_2)-\sum_{j\in S_2}\nabla Q_j(x_{\H})}\leq\mnorm{S_2}\mu\epsilon.
\end{align}
Note that $\sum_{j\in S_2}\nabla Q_j(x)=\sum_{j\in S_1}\nabla Q_j(x)+\nabla Q_i(x)$. From triangle inequality,
\begin{equation}
    \norm{\nabla Q_i(x_{\H})}-\norm{\sum_{j\in S_1}\nabla Q_j(x_{\H})}\leq\norm{\sum_{j\in S_1}\nabla Q_j(x_{\H})+\nabla Q_i(x_{\H})}.
\end{equation}
Therefore, for each non-faulty agent $i\in\H$, 
\begin{align}
    \norm{\nabla Q_i(x_{\H})}&\leq\norm{\sum_{j\in S_1}\nabla Q_j(x_{\H})+\nabla Q_i(x_{\H})}+\norm{\sum_{j\in S_1}\nabla Q_j(x_{\H})} \leq\mnorm{S_2}\mu\epsilon+\mnorm{S_1}\mu\epsilon \nonumber\\
    & = (m-2f+1)\mu \epsilon + (m-2f) \mu \epsilon = (2m-4f+1)\mu\epsilon.
    \label{eqn:honest-norm-bound}
\end{align}
Now, for all $x$ and $i\in\H$, by Assumption \ref{assum:lipschitz-ft},
\begin{equation*}
    \norm{\nabla Q_i(x)-\nabla Q_i(x_{\H})}\leq\mu\norm{x-x_{\H}}.
\end{equation*}
By triangle inequality,
\begin{equation*}
    \norm{\nabla Q_i(x)}\leq\norm{\nabla Q_i(x_{\H})}+\mu\norm{x-x_{\H}}.
\end{equation*}
Substituting from~\eqref{eqn:honest-norm-bound} above we obtain that
\begin{equation}
    \norm{\nabla Q_i(x)}\leq(2m-4f+1)\mu\epsilon+\mu\norm{x-x_{\H}}\leq2m\mu\epsilon+\mu\norm{x-x_{\H}}.
    \label{eqn:honest-bound-everywhere}
\end{equation}
We use the above inequality~\eqref{eqn:honest-bound-everywhere} to show below that $\norm{\sum_{j=1}^{m-f}g_{i_j}^t}$ is bounded for all $t$. Recall that for each iteration $t$, 
\begin{equation*}
    \norm{g_{i_1}^t}\leq...\leq\norm{g_{i_{m-f}}^t}\leq\norm{g_{i_{m-f+1}}^t}\leq...\leq\norm{g_{i_m}^t}.
\end{equation*}
As there are at most $f$ Byzantine agents, for each $t$ there exists $\sigma_t\in\H$ such that
\begin{equation}
    \norm{g_{i_{m-f}}^t}\leq\norm{g_{i_{\sigma_t}}^t}.
    \label{eqn:honest-bound}
\end{equation}
As $g_j^t=\nabla Q_j(x^t)$ for all $j\in\H$, from~\eqref{eqn:honest-bound} we obtain that
\begin{equation*}
    \norm{g_{i_j}^t}\leq\norm{\nabla Q_{\sigma_t}(x^t)}, \quad \forall j \in \{1, \ldots, m-f\}, ~ t.
\end{equation*}
Substituting from~\eqref{eqn:honest-bound-everywhere} above we obtain that for every $j \in \{1, \ldots, m-f\}$,
\begin{equation*}
    \norm{g_{i_j}^t}\leq\norm{g_{i_{n-f}}^t}\leq2m\mu\epsilon+\mu\norm{x^t-x_{\H}}.
\end{equation*}
Therefore, from triangle inequality,
\begin{equation}
    \norm{\sum_{j=1}^{m-f}g_{i_j}^t}\leq\sum_{j=1}^{m-f}\norm{g_{i_j}^t}\leq(m-f)\left(2m\mu\epsilon+\mu\norm{x^t-x_{\H}}\right). \label{eqn:filtered-upperbound}
\end{equation}
Recall from~\eqref{eqn:x_H_W} that $x_{\H} \in \W$. Let $\Gamma = \max_{x \in \W} \norm{x - x_{\H}}$. As $\W$ is a compact set, $\Gamma < \infty$. Recall from the update rule~\eqref{eqn:update} that $x^t \in \W$ for all $t$. Thus, $\norm{x^t - x_{\H}} \leq \max_{x \in \W} \norm{x - x_{\H}} = \Gamma < \infty$. Substituting this in~\eqref{eqn:filtered-upperbound} implies that
\begin{equation}
    \norm{\sum_{j=1}^{m-f}g_{i_j}^t} \leq (m-f) \left( 2m \mu \epsilon + \mu \Gamma\right) < \infty. 
\end{equation}
Recall that in this particular case, $\sum_{j=1}^{m-f}g_{i_j}^t = \gf\left(m,f;\,\left\{g^t_i\right\}_{i\in S^t} \right)$ (see \eqref{eqn:cge_gf}). Therefore, from above we obtain that
\begin{align}
    \norm{\gf\left(m,f;\,\left\{g^t_i\right\}_{i\in S^t} \right)} < \infty, \quad \forall t. \label{eqn:cge_bnd_grd}
\end{align}

\textbf{Second}, we show that for an arbitrary $\delta > 0$ there exists $\xi > 0$ such that
\[\phi_t\triangleq\iprod{x^t-x_{\H}}{\sum_{j=1}^{m-f}g_{i_j}^t} \geq \xi ~ \text{ when } ~ \norm{x^t-x_{\H}} \geq \D \, \epsilon + \delta.\]
Consider an arbitrary iteration $t$. Note that, as $\mnorm{\H} = n-f$, there are at least $m-2f$ agents that are common to both sets $\H$ and $\{i_1,...,i_{m-f}\}$. We let $\H^t = \{i_1,...,i_{m-f}\} \cap \H$. The remaining set of agents $\B^t = \{i_1,...,i_{m-f}\} \setminus \H^t$ comprises of only faulty agents. Note that $\mnorm{\H^t} \geq m-2f $ and $\mnorm{\B^t} \leq f$. Therefore,
\begin{equation}
    \phi_t=\iprod{x^t-x_{\H}}{\sum_{j\in\H^t}g_j^t}+\iprod{x^t-x_{\H}}{\sum_{k\in\B^t}g_k^t}.
    \label{eqn:phi-t-two-parts}
\end{equation}
Consider the first term in the right-hand side of~\eqref{eqn:phi-t-two-parts}. Note that
\begin{align*}
    \iprod{x^t-x_{\H}}{\sum_{j\in\H^t}g_j^t}&=\iprod{x^t-x_{\H}}{\sum_{j\in\H^t}g_j^t+\sum_{j\in\H\backslash\H^t}g_j^t-\sum_{j\in\H\backslash\H^t}g_j^t} \\ 
    &=\iprod{x^t-x_{\H}}{\sum_{j\in\H}g_j^t}-\iprod{x^t-x_{\H}}{\sum_{j\in\H\backslash\H^t}g_j^t}.
\end{align*}
Recall that $g_j^t=\nabla Q_j(x^t)$, $\forall j\in\H$. Therefore,
\begin{equation}
    \iprod{x^t-x_{\H}}{\sum_{j\in\H^t}g_j^t}=\iprod{x^t-x_{\H}}{\sum_{j\in\H}\nabla Q_j(x^t)}-\iprod{x^t-x_{\H}}{\sum_{j\in\H\backslash\H^t}\nabla Q_j(x^t)}. \label{eqn:first_phi_1}
\end{equation}

\noindent Due to the strong convexity assumption (i.e., Assumption~\ref{assum:strongly-convex-ft}), for all $x, \, y \in \R^d$,
\begin{equation*}
    \iprod{x- y}{\nabla \sum_{j\in\H}Q_j(x)-\nabla\sum_{j\in\H} Q_j(y)} \geq \mnorm{\H}\, \gamma\norm{x-y}^2.
\end{equation*}
As $x_{\H}$ is minimum point of $\sum_{j \in \H}Q_j(x)$, $\nabla \sum_{j \in \H} Q_j(x_{\H}) = 0$. Thus, 
\begin{align}
    \iprod{x^t-x_{\H}}{\sum_{j\in\H}\nabla Q_j(x^t)} & = \iprod{x^t-x_{\H}}{\nabla\sum_{j\in\H}Q_j(x^t)-\nabla\sum_{j\in\H}Q_j(x_{\H})} \nonumber \\
    & \geq \mnorm{\H} \, \gamma\norm{x^t-x_{\H}}^2.
    \label{eqn:inner-prod-h}
\end{align}
Now, due to the Cauchy-Schwartz inequality, 
\begin{align}
    \iprod{x^t-x_{\H}}{\sum_{j\in\H\backslash\H^t}\nabla Q_j(x^t)}&=\sum_{j\in\H\backslash\H^t}\iprod{x^t-x_{\H}}{\nabla Q_j(x^t)} \nonumber\\
    &\leq\sum_{j\in\H\backslash\H^t}\norm{x^t-x_{\H}}\, \norm{\nabla Q_j(x^t)}.
    \label{eqn:inner-prod-h-ht}
\end{align}
Substituting from~\eqref{eqn:inner-prod-h} and~\eqref{eqn:inner-prod-h-ht} in~\eqref{eqn:first_phi_1} we obtain that
\begin{equation}
    \iprod{x^t-x_{\H}}{\sum_{j\in\H^t}g_j^t} \geq \gamma\mnorm{\H}\, \norm{x^t-x_{\H}}^2-\sum_{j\in\H\backslash\H^t}\norm{x^t-x_{\H}}\, \norm{\nabla Q_j(x^t)}.
    \label{eqn:phi-t-first-part}
\end{equation}

Next, we consider the second term in the right-hand side of~\eqref{eqn:phi-t-two-parts}. From the Cauchy-Schwartz inequality, 
\begin{equation*}
    \iprod{x^t-x_{\H}}{g_k^t}\geq-\norm{x^t-x_{\H}}\, \norm{g_k^t}.
\end{equation*}
Substituting from~\eqref{eqn:phi-t-first-part} and above in~\eqref{eqn:phi-t-two-parts} we obtain that
\begin{equation}
    \phi_t\geq\gamma\mnorm{\H}\, \norm{x^t-x_{\H}}^2-\sum_{j\in\H\backslash\H^t}\norm{x^t-x_{\H}}\, \norm{\nabla Q_j(x^t)}-\sum_{k\in\B^t}\norm{x^t-x_{\H}}\, \norm{g_k^t}.
    \label{eqn:phi-t-2}
\end{equation}

Recall that, due to the sorting of the gradients, for an arbitrary $k \in \B^t$ and an arbitrary $j \in \H\backslash\H^t$,
\begin{equation}
    \norm{g_k^t}\leq\norm{g_j^t}=\norm{\nabla Q_j(x^t)}. \label{eqn:k_B_j_H}
\end{equation}
Recall that $\B^t = \{i_1, \ldots, \, i_{m-f}\} \setminus \H^t$. Thus, $\mnorm{\B^t} = m-f-\mnorm{\H^t}$. Also, as $\mnorm{\H} = n-f$, $\mnorm{\H\backslash\H^t} = n- f - \mnorm{\H^t}$. That is, $\mnorm{\B^t} \leq \mnorm{\H\backslash\H^t}$. Therefore,~\eqref{eqn:k_B_j_H} implies that
\begin{align*}
    \sum_{k \in \B^t} \norm{g_k^t} \leq \sum_{j \in \H\backslash\H^t} \norm{\nabla Q_j(x^t)}.
\end{align*}
Substituting from above in~\eqref{eqn:phi-t-2}, we obtain that
\begin{align*}
    \phi_t&\geq\gamma\mnorm{\H}\, \norm{x^t-x_{\H}}^2-2\sum_{j\in\H\backslash\H^t}\norm{x^t-x_{\H}}\, \norm{\nabla Q_j(x^t)}.
\end{align*}
Substituting from \eqref{eqn:honest-bound-everywhere}, i.e., $\norm{\nabla Q_i(x)}\leq2m\mu\epsilon+\mu\norm{x-x_{\H}}$, above we obtain that
\begin{align*}
    \phi_t&\geq\gamma\mnorm{\H}\, \norm{x^t-x_{\H}}^2 - 2\mnorm{\H\backslash\H^t}\, \norm{x^t-x_{\H}}\, (2m\mu\epsilon+\mu\norm{x^t-x_{\H}}) \nonumber\\
    & \geq \left(\gamma\mnorm{\H}-2\mu\mnorm{\H\backslash\H^t}\right)\norm{x^t-x_{\H}}^2 - 4m\mu\epsilon \mnorm{\H\backslash\H^t}\, \norm{x^t-x_{\H}}.
\end{align*}
As $\mnorm{\H} = n-f$ and $\mnorm{\H\backslash\H^t}\leq f+r$, the above implies that
\begin{align}
    \begin{split}
    \label{eqn:phi-t-pre-final}
        \phi_t & \geq\left(\gamma(n-f)-2\mu (f+r)\right)\norm{x^t-x_{\H}}^2-4m\mu\epsilon (f+r) \norm{x^t-x_{\H}} \\
            & = \left(\gamma(n-f)-2\mu (f+r)\right)\norm{x^t-x_{\H}} \left(\norm{x^t-x_{\H}}-\dfrac{4m\mu\epsilon (f+r)}{\gamma(n-f)-2\mu (f+r)} \right) \\
            & = m \gamma \left( 1 - \frac{f-r}{m} + \frac{2\mu}{\gamma}\cdot\frac{f+r}{m}\right) \norm{x^t-x_{\H}}  \left(\norm{x^t-x_{\H}} - \frac{4 \mu (f+r) \epsilon}{\gamma\left( 1 - \frac{f-r}{m} + \frac{2\mu}{\gamma}\cdot\frac{f+r}{m}\right)} \right).
    \end{split}
\end{align}
Recall that we defined 
\[\alpha = 1 - \frac{f-r}{m} + \frac{2\mu}{\gamma}\cdot\frac{f+r}{m}. \]
Substituting from above in~\eqref{eqn:phi-t-pre-final} we obtain that
\begin{align}
    \phi_t \geq \alpha m \gamma \norm{x^t-x_{\H}} \left(\norm{x^t-x_{\H}}- \frac{4\mu (f+r)\epsilon}{\alpha\gamma} \right).
    \label{eqn:phi-t-final}
\end{align}
As it is assumed that $\alpha > 0$,~\eqref{eqn:phi-t-final} implies that for an arbitrary $\delta > 0$,
\begin{align}
    \phi_t \geq \alpha m \gamma \delta \left( \frac{4\mu (f+r)\epsilon}{\alpha\gamma} + \delta \right) > 0 ~ \text{ when } \norm{x^t-x_{\H}}>\frac{4\mu (f+r)\epsilon}{\alpha\gamma}.
    \label{eqn:final-cge}
\end{align}
Combine \eqref{eqn:final-cge} with Theorem~\ref{thm:async-fault-toler}, hence, the proof.